\documentclass[12pt,a4paper]{article}

\usepackage{amsmath,amssymb,slashed,cancel}
\usepackage{cite,tabularx}
\usepackage{subcaption}
\usepackage{graphicx,color}
\usepackage[normalem]{ulem}
\usepackage{mathtools}
\usepackage{enumerate}
\allowdisplaybreaks

\usepackage[height=8.85in,width=6.4in]{geometry}
\renewcommand{\baselinestretch}{1.1}
\newcommand\package[2][\relax]{\texttt{#2\ifx#1\relax\relax\relax\else\,\linebreak[0]#1\fi}}

\numberwithin{equation}{section} 
 
\def\beq#1\eeq{\begin{align}#1\end{align}}

\definecolor{BlueViolet}{rgb}{0.2, 0.00, 0.7}
\definecolor{Blue}{rgb}{0.15, 0.00, 0.9}
\usepackage[colorlinks=true,linkcolor=Blue,citecolor=Blue,urlcolor=BlueViolet]{hyperref}

\begin{document}
\begin{titlepage}
\setcounter{page}{0} 

\begin{center}

\vskip .55in

\begingroup
\centering
\large\bf SMEFT effects on gravitational wave spectrum from electroweak phase transition
\endgroup

\vskip .4in

{
  Katsuya Hashino$^{\rm (a)}$, and
  Daiki Ueda$^{\rm (b)}$
}

\vskip 0.4in

\begingroup\small
\begin{minipage}[t]{0.9\textwidth}
\centering\renewcommand{\arraystretch}{0.9}
\begin{tabular}{c@{\,}l}
$^{\rm(a)}$
& Department of Physics, Faculty of Science and Technology,
Tokyo University of Science,\\
&Noda, Chiba 278-8510, Japan \\[2mm]
$^{\rm (b)}$
& Center for High Energy Physics, Peking University, Beijing 100871, China\\
\end{tabular}
\end{minipage}
\endgroup

\end{center}

\vskip .4in

\begin{abstract}\noindent
Future gravitational wave observations are potentially sensitive to new physics corrections to the Higgs potential once the first-order electroweak phase transition arises. 
We study the SMEFT dimension-six operator effects on the Higgs potential, where three types of effects are taken into account: (i) SMEFT tree level effect on $\varphi^6$ operator, (ii) SMEFT tree level effect on the wave function renormalization of the Higgs field, and (iii) SMEFT top-quark one-loop level effect.
The sensitivity of future gravitational wave observations to these effects is numerically calculated by performing a Fisher matrix analysis.
We find that the future gravitational wave observations can be sensitive to (ii) and (iii) once the first-order electroweak phase transition arises from (i).
The dimension-eight $\varphi^8$ operator effects on the first-order electroweak phase transition are also discussed.
The sensitivities of the future gravitational wave observations are also compared with those of future collider experiments. 
\end{abstract}
\end{titlepage}

\setcounter{page}{1}
\renewcommand{\thefootnote}{\#\arabic{footnote}}
\setcounter{footnote}{0}

\begingroup
\renewcommand{\baselinestretch}{1} 
\setlength{\parskip}{2pt}          
\hrule
\tableofcontents
\vskip .2in
\hrule
\vskip .4in
\endgroup

\section{Introduction}\label{sec:intro}

The CERN large hadron collider (LHC) has discovered the Higgs boson~\cite{ATLAS:2012yve,CMS:2012qbp} and measured its properties closely resembling the Standard Model (SM).
The discovery of the Higgs boson strengthened the conviction of the SM. 
However, the shape of the Higgs potential is still unknown, and determining the nature of the electroweak phase transition would be a major scientific goal. %
In particular, the strongly first-order electroweak phase transition (SFO-EWPT) could provide suitable conditions for achieving the observed baryon asymmetry of the universe (BAU)~\cite{Gavela:1993ts,Konstandin:2003dx,Kuzmin:1985mm} in 
the electroweak baryogenesis scenario, but the SFO-EWPT in the SM only arises for a Higgs mass $m_h\lesssim 65$ GeV~\cite{Kajantie:1995kf,Kajantie:1996mn,Kajantie:1996qd,Csikor:1998eu,DOnofrio:2015gop} well below the measured Higgs mass $125$ GeV~\cite{Workman:2022ynf}.
In these circumstances, there is growing attention to new physics~(NP) effects on the Higgs potential from both theoretical and experimental points of view, but the lack of new particle discoveries at the LHC strengthens the possibility of the NP scale higher than the electroweak symmetry breaking (EWSB) scale.

This situation motivates the effective field theory (EFT) approach to describe the NP effects.
The Standard Model Effective Field Theory (SMEFT)~\cite{Grzadkowski:2010es,Jenkins:2013zja,Jenkins:2013wua,Alonso:2013hga} is one of the actively studied EFTs, and information about the NP effects is transferred to higher-dimensional operators of EFTs consisting of the SM fields.
To generate the SFO-EWPT, a considerable amount of literature~\cite{Grojean:2004xa,Zhang:1992fs,Bodeker:2004ws,Huber:2007vva,Delaunay:2007wb,Huber:2013kj,Konstandin:2014zta,Damgaard:2015con,Harman:2015gif,Balazs:2016yvi,deVries:2017ncy,Cai:2017tmh,Chala:2018ari,Dorsch:2018pat,DeVries:2018aul,Chala:2019rfk,Ellis:2019flb,Zhou:2019uzq,Kanemura:2020yyr,Phong:2020ybr,Wang:2020zlf,Wang:2020jrd,Kanemura:2021fvp,Lewicki:2021pgr,Hashino:2021qoq,Kanemura:2022txx,Anisha:2022hgv,Croon:2020cgk,Huang:2015izx,Cao:2017oez,Huang:2016odd} has considered the SMEFT dimension-six $\varphi^6$ operator\footnote{In Refs.~\cite{Damgaard:2015con,Postma:2020toi}, the validity of the SMEFT description for the SFO-EWPT is questioned; see Sec.~\ref{sec:Dis}.}.
In the context of the electroweak baryogenesis scenario, the other SMEFT dimension-six operators~\cite{Bodeker:2004ws,Balazs:2016yvi,Huber:2006ri,Huang:2015izx,Cao:2017oez,Huang:2016odd} are also studied.
On the experimental grounds, there is growing interest in the constraints on the SMEFT Wilson coefficients from the current and past experimental data, and future collider experiments, e.g., high-luminosity LHC~\cite{Cepeda:2019klc}, the International Linear Collider (ILC)~\cite{LCCPhysicsWorkingGroup:2019fvj}, the Compact LInear Collider (CLIC)~\cite{CLIC:2018fvx}, the Future Circular Collider of electrons and positrons (FCC-ee)~\cite{FCC:2018byv}, and the Circular Electron Positron Collider (CEPC)~\cite{An:2018dwb}. 
Furthermore, the SFO-EWPT predicts stochastic background of gravitational waves (GWs), and its spectrum can be peaked around the future interferometer experiment band with milli- to deci-Hertz, such as Laser Interferometer Space Antenna (LISA)~\cite{LISA:2017pwj}, DECi-hertz Interferometer Gravitational wave Observatory (DECIGO)~\cite{Seto:2005qy}, and Big-Bang Observer (BBO)~\cite{BBO}.
Therefore, the sensitivities of future GW observations to the SMEFT $\varphi^6$ operator also have been investigated~\cite{Huber:2007vva,Delaunay:2007wb,Cai:2017tmh,Chala:2018ari,Chala:2019rfk,Ellis:2019flb,Zhou:2019uzq,Wang:2020jrd,Lewicki:2021pgr,Hashino:2021qoq,Kanemura:2022txx,Croon:2020cgk,Huang:2016odd}.

The previous works mainly studied a parameter space to generate a detectable amount of GWs, but they have not quantified how precisely the NP effects can be measured once the GWs are detected.
In light of these circumstances, in the previous works of the NP search by the GW observations~\cite{Hashino:2018wee}, the method of Fisher matrix analysis was proposed to evaluate the expected sensitivities to NP model parameters.
This analysis quantifies how precisely the NP model parameters can be measured by the GWs observations, and it is clarified that the GW observations potentially have higher sensitivities to small deviations of the Higgs potential by the NP effects than the future collider experiments such as the ILC-250.
This result naturally leads us to study the sub-dominant SMEFT effects on the Higgs potential and the sensitivities of the GW observations to them.

In this paper, we study the SMEFT dimension-six operator corrections to the Higgs potential and the sensitivities of the GW observations to them.
We will focus on three types of the SMEFT dimension-six operator effects: (i) SMEFT tree level effect on $\varphi^6$, (ii) SMEFT tree level effects on the wave function renormalization of the Higgs field, and (iii) SMEFT one-loop top-quark effects.
Type (i) dominates the SMEFT effect on the Higgs potential and can achieve the SFO-EWPT.
Type (ii) is the tree level effects, but not dominant effects because of the suppression by interference effects with the Higgs self couplings. 
Type (iii) arises from the loop diagrams and can not dominate the SMEFT effect, but it can be a measurable effect because of the large top Yukawa coupling. 
Therefore, we focus on a scenario where the SFO-EWPT mainly arises by (i), and the Higgs potential is slightly shifted by (ii) and (iii).
We will evaluate the SMEFT effects on the GW spectrum and perform the Fisher matrix analysis to clarify the expected sensitivities of future GW observations to (ii) and (iii).
Their expected sensitivities to (ii) and (iii) are also discussed when the SMEFT dimension-eight $\varphi^8$ operator is added.

This paper is organized as follows.
In Sec.~\ref{sec:form}, we provide formulae of the SMEFT dimension-six operator effects on the Higgs potential, and in the following section, we evaluate the SMEFT effects on the SFO-EWPT.
In Sec.~\ref{sec:GWsp}, we briefly review Refs.~\cite{Caprini:2019egz,Hindmarsh:2017gnf,PhysRevD.101.089902,Espinosa:2010hh,Seto:2005qy,Hashino:2018wee,Yagi:2011wg,Klein:2015hvg} and summarize formulae of the GW spectrum from the SFO-EWPT and how to evaluate the sensitivities of future GWs observations to the SMEFT effects, e.g., the Fisher matrix analysis.
In Sec.~\ref{sec:res}, the results of numerical calculations are collected. 
In Sec.~\ref{sec:Dis}, the SMEFT dimension-eight operator effects on the SFO-EWPT are discussed.
We finish with the summary of the paper in Sec.~\ref{sec:summ}.

\section{Formula}\label{sec:form}
The information of the NP particles is transferred to higher-dimensional operators of the SMEFT when the NP scale is higher than the EWSB scale.
The SMEFT operators contribute to the Higgs potential and affect the EWPT. 
In this section, we provide the formulae of the SMEFT dimension-six operator effects on the Higgs potential by taking into account the SMEFT effects on the wave function renormalization of the Higgs field and the SMEFT top-quark effects.
In particular, we newly consider the dimension-six $\mathcal{O}_{uH}$ operator effects.
For convenience, we also summarize the $\mathcal{O}_{H}$, $\mathcal{O}_{HD}$, and $\mathcal{O}_{H\Box}$ effects as studied in Refs.~\cite{Zhang:1992fs,Grojean:2004xa,Bodeker:2004ws,Huber:2007vva,Delaunay:2007wb,Huber:2013kj,Konstandin:2014zta,Damgaard:2015con,Harman:2015gif,deVries:2017ncy,Cai:2017tmh,Dorsch:2018pat,Chala:2019rfk,Ellis:2019flb,Zhou:2019uzq,Kanemura:2020yyr,Phong:2020ybr,Wang:2020zlf,Wang:2020jrd,Lewicki:2021pgr,Anisha:2022hgv,Croon:2020cgk,Huang:2015izx,Cao:2017oez,Huang:2016odd,Postma:2020toi}.
The Lagrangian of the SMEFT is defined as~\cite{Grzadkowski:2010es}
\begin{align}
    \mathcal{L}_{\rm SMEFT}=\mathcal{L}_{\rm SM}+\sum_i C_i \mathcal{O}_i,\label{eq:LSMEFT}
\end{align}
where the first term on the right-hand side is the SM Lagrangian, and the second term denotes the higher-dimensional operators consisting of the SM fields.
The Lagrangian of Eq.~\eqref{eq:LSMEFT} is invariant under the SM gauge symmetry, and all the SM particles, e.g., $W, Z, H$, and $t$, are dynamical.
We consider the SMEFT operators involving Higgs and top-quarks, which contribute to the Higgs potential since the top Yukawa coupling is large.  
For simplicity, we will restrict this study to CP-conserving interactions.
The dimension-six operators relevant to the Higgs potential up to the tree level are
\begin{align}
        &\mathcal{O}_{H\Box}=(H^{\dagger}H)\Box (H^{\dagger}H),
    \\
    &\mathcal{O}_{HD}=(H^{\dagger}D^{\mu}H)^{\ast}(H^{\dagger}D_{\mu}H),
    \\
    &\mathcal{O}_{H}=(H^{\dagger}H)^3,
\end{align}
where the Higgs field is written in unitary gauge as $\sqrt{2}H^T =(0,\varphi)=(0,v+h)$ with $v=246~{\rm GeV}$.
The dimension-six operators involving the Higgs fields and top quarks are listed as follows,
\begin{align}
    &(\mathcal{O}_{uH})_{ij}=(H^{\dagger}H)(\bar{q}_i u_j \tilde{H}),
    \\
    &(\mathcal{O}_{Hq}^{(1)})_{ij}=(H^{\dagger}i\overleftrightarrow{D}_{\mu}H)(\bar{q}_i \gamma^{\mu}q_j),\label{eq:Hq1}
    \\
    &(\mathcal{O}_{Hq}^{(3)})_{ij}=(H^{\dagger}i\overleftrightarrow{D}_{\mu}^IH)(\bar{q}_i \tau^I \gamma^{\mu}q_j),\label{eq:Hq3}
    \\
    &(\mathcal{O}_{Hu})_{ij}=(H^{\dagger}i\overleftrightarrow{D}_{\mu}H)(\bar{u}_i  \gamma^{\mu}u_j),\label{eq:Hu}
\end{align}
with the derivative
\begin{align}
    H^{\dagger}\overleftrightarrow{D}^I_{\mu} H=H^{\dagger} \tau^I D_{\mu} H-(D_{\mu}H)^{\dagger}\tau^I H.
\end{align}
Here, $q$ is the $SU(2)_L$ quark doublet, $u$ the right-handed up-type quark, quark-flavor indices $i,j$, an $SU(2)_L$ index $I$, and $\tau^I$ the Pauli matrices.
The Higgs Lagrangian including the SMEFT corrections is defined as
\begin{align}
    \mathcal{L}_{\varphi}=\frac{1}{2}(\partial_{\mu}\varphi)^2-\frac{1}{2}\mu^2 \varphi^2-\frac{1}{4}\lambda \varphi^4+\Delta \mathcal{L}_{\rm SMEFT},\label{eq:higgLag}
\end{align}
where the first, second, and third terms represent the SM renormalizable interactions, and the last term denotes the SMEFT corrections. 
We summarize the SMEFT corrections to the Higgs Lagrangian for each operator as follows:
\begin{itemize}
    \item $\mathcal{O}_{H}$ ---
    Substituting the Higgs field $\sqrt{2} H^T =(0,\varphi)$ into $\mathcal{O}_{H}$ yields the correction at the tree-level as follows,
    \begin{align}
        \Delta \mathcal{L}_{\rm SMEFT}=\frac{1}{8} C_H \varphi^6.\label{eq:SMEFTCH}
    \end{align}
    As studied in Ref.~\cite{Zhang:1992fs,Grojean:2004xa,Bodeker:2004ws,Huber:2007vva,Delaunay:2007wb,Huber:2013kj,Konstandin:2014zta,Damgaard:2015con,Harman:2015gif,deVries:2017ncy,Cai:2017tmh,Dorsch:2018pat,Chala:2019rfk,Ellis:2019flb,Zhou:2019uzq,Kanemura:2020yyr,Phong:2020ybr,Wang:2020zlf,Wang:2020jrd,Lewicki:2021pgr,Anisha:2022hgv,Croon:2020cgk,Huang:2015izx,Cao:2017oez,Huang:2016odd}, the $\varphi^6$ operator can give rise to the SFO-EWPT and is a dominant SMEFT effect on the Higgs Lagrangian.
    Although, as studied in Refs.~\cite{Damgaard:2015con,Postma:2020toi}, the SMEFT dimension-six operator descriptions of the SFO-EWPT are limited, the SMEFT with an additional dimension-eight $\varphi^8$ operator can be UV completed 
    in an extended model with a singlet scalar boson and describe the SFO-EWPT~\cite{Postma:2020toi}.
    The additional dimension-eight operator effects on the sensitivity reach of the future GW observations are discussed in Sec.~\ref{sec:Dis}.
    \item $\mathcal{O}_{HD}$ ---
    As shown in Ref.~\cite{Postma:2020toi}, this operator yields the correction at the tree level to the Higgs Lagrangian as follows,
    \begin{align}
        \Delta \mathcal{L}_{\rm SMEFT}=\frac{1}{4}C_{HD} \varphi^2 (\partial_{\mu}\varphi)^2.\label{eq:CH}
    \end{align}
    As discussed in the next section, Eq.~\eqref{eq:CH} contributes to the Higgs potential by the wave function renormalization of $\varphi$. 
    
    \item $\mathcal{O}_{H\Box}$ ---
    As shown in Ref.~\cite{Postma:2020toi}, this operator yields the correction at the tree level to the Higgs Lagrangian as follows,
    \begin{align}
        \Delta \mathcal{L}_{\rm SMEFT}=-C_{H\Box} \varphi^2 (\partial_{\mu}\varphi)^2.\label{eq:CBox}
    \end{align}
    Similar to $\mathcal{O}_{HD}$,  Eq.~\eqref{eq:CBox} contributes to the Higgs potential by the wave function renormalization of $\varphi$. 
    
    \item $\mathcal{O}_{uH}$ ---
    We newly consider $\mathcal{O}_{uH}$ operator effects on the Higgs Lagrangian.
    By the top-quark one-loop effects, $\mathcal{O}_{uH}$ generates the corrections to the Higgs Lagrangian as follows,
    \begin{align}
        \Delta \mathcal{L}_{\rm SMEFT}&=C_{uH}\cdot \frac{3}{32\pi^2}Y_t  \left(14-6\ln \frac{m_t^2}{v^2}\right)\cdot\frac{1}{2}\varphi^2(\partial_{\mu}\varphi)^2-\Delta V_{c_{uH}},\label{eq:CuHloop}
    \end{align}
    where $Y_t$ is the top Yukawa coupling, $m_t$ is the mass of the top-quark, and
    \begin{align}
        \Delta V_{c_{uH}}&\equiv - \frac{12}{64\pi^2}\bigg[
        m^4_t(\varphi,C_{uH}) \left(\ln \frac{m^2_t(\varphi,C_{uH})}{v^2} -\frac{3}{2}\right)
        -m^4_t(\varphi,0) \left(\ln \frac{m^2_t(\varphi,0)}{v^2} -\frac{3}{2}\right)\bigg]\label{eq:int},
        \\
        &=-C_{uH}\cdot \frac{3}{32\pi^2} Y_t^3 \varphi^6 \left(-1+\ln \frac{Y_t^2 \varphi^2}{2 v^2}\right)+\mathcal{O}(C_{uH}^2).\label{eq:CWfist}
    \end{align}
    with
    \begin{align}
        m^2_t(\varphi,C_{uH})\equiv \left(\frac{Y_t \varphi}{\sqrt{2}}+\frac{C_{uH} \varphi^3}{2\sqrt{2}}\right)^2.
    \end{align}
    Here, the $\overline{\rm MS}$ regularization scheme is adopted, and the results are evaluated at the EWSB scale $v=246~{\rm GeV}$.
    The first term of Eq.~\eqref{eq:CuHloop}, and Eq.~\eqref{eq:CWfist} are obtained by the fermionic universal one-loop effective action (UOLEA)~\cite{Ellis:2020ivx} calculated by the covariant derivative expansion, and they are consistent with the one-loop anomalous dimension matrix for the dimension-six operators of the SMEFT~\cite{Jenkins:2013wua}.
    As shown in Eq.~\eqref{eq:int} and \eqref{eq:CWfist}, we checked that the one-loop Coleman-Weinberg~(CW) potential~\cite{Coleman:1973jx} with $C_{uH}$ effect is consistent with the results of the UOLEA up to the first order of $C_{uH}$. 
    The first term of Eq.~\eqref{eq:CuHloop} contributes to the Higgs potential by the wave function renormalization of $\varphi$.

    \item $\mathcal{O}_{Hq}^{(1)}$, $\mathcal{O}_{Hq}^{(3)}$, $\mathcal{O}_{Hu}$ ---
    These operators potentially involve both Higgs fields and top-quark.
    Since, however, the neutral Higgs field in the derivative of Eq.~\eqref{eq:Hq1}, \eqref{eq:Hq3}, and \eqref{eq:Hu} cancels, the SMEFT corrections to the Higgs potential can not arise. 
\end{itemize}

\section{First-order phase transitions in SMEFT}
\label{sec:Fr}
We consider the EWPT described by the Higgs potential involving the SMEFT effects of Sec.~\ref{sec:form} and discuss the conditions for achieving the SFO-EWPT.
As explained in Sec.~\ref{sec:form}, we newly consider the dimension-six $\mathcal{O}_{uH}$ operator effects and summarize the $\mathcal{O}_H$, $\mathcal{O}_{HD}$, and $\mathcal{O}_{H\Box}$ effects as studied in Refs.~\cite{Zhang:1992fs,Grojean:2004xa,Bodeker:2004ws,Huber:2007vva,Delaunay:2007wb,Huber:2013kj,Konstandin:2014zta,Damgaard:2015con,Harman:2015gif,deVries:2017ncy,Cai:2017tmh,Dorsch:2018pat,Chala:2019rfk,Ellis:2019flb,Zhou:2019uzq,Kanemura:2020yyr,Phong:2020ybr,Wang:2020zlf,Wang:2020jrd,Lewicki:2021pgr,Anisha:2022hgv,Croon:2020cgk,Huang:2015izx,Cao:2017oez,Huang:2016odd,Postma:2020toi}.

\subsection{Higgs potential}
We summarize the SMEFT effects on the Higgs Lagrangian at zero temperature.
Combining Eq.~\eqref{eq:higgLag}, \eqref{eq:SMEFTCH}, \eqref{eq:CH}, \eqref{eq:CBox}, and \eqref{eq:CuHloop}, we obtain the following Lagrangian:
\begin{align}
    \mathcal{L}_{\varphi}&=c_{\rm kin}\varphi^2 (\partial_{\mu}\varphi)^2+\frac{1}{8} C_H \varphi^6
    -\Delta V_{c_{uH}},\label{eq:smeftvpo}
\end{align}
Here, $c_{\rm kin}$ is defined as
\begin{align}
    c_{\rm kin}&=c_{\rm kin}^{(0)}+c_{\rm kin}^{(1)}\cdot \varphi+c_{\rm kin}^{(2)}\cdot \varphi^2,
\end{align}
with
\begin{align}
\label{ckin0}
    &c_{\rm kin}^{(0)}=\frac{1}{4}C_{HD}-C_{H\Box}+\frac{1}{2}C_{uH}\cdot \frac{3}{32\pi^2}Y_t  \left(14-6\ln \frac{m_t^2}{v^2}\right),
    \\
\label{ckin1}
    &c_{\rm kin}^{(1)}=\frac{1}{2}C_{uH}\cdot \frac{3}{32\pi^2}Y_t  \left(-\frac{28}{v}\right),
    \\
\label{ckin2}
    &c_{\rm kin}^{(2)}=\frac{1}{2}C_{uH}\cdot \frac{3}{32\pi^2}Y_t  \left(\frac{8}{v^2}\right).
\end{align}
By a field redefinition,
\begin{align}
 \varphi\to \varphi-\frac{1}{3} c_{\rm kin}^{(0)}\cdot \varphi^3-\frac{1}{4} c_{\rm kin}^{(1)}\cdot \varphi^4-\frac{1}{5} c_{\rm kin}^{(2)}\cdot \varphi^5 +\mathcal{O}(c_{\rm kin}^2),
\end{align} 
we obtain the Higgs Lagrangian as
\begin{align}
    \mathcal{L}_{\varphi}=\frac{1}{2}(\partial_{\mu}\varphi)^2 -V,
\end{align}
with the Higgs potential up to the first order of the Wilson coefficients,
\begin{align}
    V=\frac{1}{2}\mu^2 \varphi^2 +\frac{1}{4}\left(\lambda -\frac{4}{3}c^{(0)}_{\rm kin}\mu^2\right)\varphi^4 -\frac{1}{4} c^{(1)}_{\rm kin} \mu^2 \varphi^5 +\frac{1}{6}\left(-\frac{3}{4}C_H -2 c^{(0)}_{\rm kin}\lambda-\frac{6}{5}c^{(2)}_{\rm kin}\mu^2 \right)\varphi^6 +\Delta V_{c_{uH}}.\label{eq:HiggsP}
\end{align}
As explained in the next section, in the numerical calculations, the one-loop CW contribution~\cite{Coleman:1973jx} and the thermal corrections of the SM particles are included in Eq.~\eqref{eq:HiggsP}.

\subsection{First-order electroweak phase transition}
The EWPT is described by the effective potential with the CW contributions and finite temperature effects:
\begin{align}
V_{\rm eff}\left(\varphi, T\right) = V + V_{\rm one\text{-}\rm loop} +  \Delta V_T + V_T^{\rm ring},\label{eq:veffnum}
\end{align}
where $V$ is defined in Eq.~\eqref{eq:HiggsP}.
%
The CW contributions in the effective potential are given by~\cite{Coleman:1973jx}, 
\begin{align}
  V_{\rm one\text{-}\rm loop} 
  = \sum_i \frac{n_i}{64\pi^2} M_i^4(\varphi) 
  \left( \ln \frac{M_i^2(\varphi)}{v^2} - c_i \right)\ , 
  \label{eq:v1}
\end{align}
%
where the $\overline{\rm MS}$ regularization scheme is adopted, the right-hand side of Eq.~\eqref{eq:v1} is evaluated at the EWSB scale $v=246$ GeV.
Here, the index $i$ runs over the Higgs, $W$, $Z$, and $t$, $c_i=5/6$ ($3/2$) for the gauge boson (other particles), and $n_i$ is the degrees of freedom of particles ($n_{\varphi} = 1, n_{W}=6, n_Z=3, n_t=-12$).
The field-dependent masses are defined as
\begin{align}
    &M^2_{\varphi}(\varphi)\equiv \mu^2+3 \lambda \varphi^2,
    \\
    &M_i^{2 {\rm (L, T)}}(\varphi)\equiv \frac{\varphi^2}{4} 
\left(\begin{array}{cccccccc} 
g^2 &0&0&0 \\
0&g^2 &0&0 \\
0&0&g^2&g g' \\
0&0&g'g&{g'}^2
\end{array}\right),~~{\rm for}~i=W^1, W^2, W^3, B,
\\
&M^2_t(\varphi)\equiv m^2_t (\varphi,0)= \frac{Y^2_t}{2}\varphi^2,
\end{align}
where superscript $\rm L$ and $\rm T$ represents longitudinal and transverse parts of the gauge bosons, and $g'$ and $g$ are the SM $U(1)$ and $SU(2)$ gauge coupling, respectively.
The parameters $\mu^2$ and $\lambda$ in the Higgs potential are fixed by the following conditions,
\begin{align}
    &\partial_{\varphi} V_{\rm eff}\left(\varphi, 0\right)|_{\varphi=v} =0,\label{eq:Vpr}
    \\
    &\partial_{\varphi}^2 V_{\rm eff}\left(\varphi, 0\right)|_{\varphi=v} =m_{h}^2,\label{eq:Vprpr}
\end{align}
where the Higgs mass is $m_{h}=125~{\rm GeV}$.
The thermal loop effects in the effective potential~\cite{PhysRevD.9.3320} are expressed as
	\begin{align}
	\label{FINI}
 \Delta V_T&= \frac{T^4}{2\pi^2} 
  \Bigl\{ \sum_{i=\varphi,W,Z} 
  n_i  \cdot I_{\rm B}\left((M_i(\varphi)/T)^2 \right) + n_t  \cdot I_{\rm F}\left((M_t(\varphi)/T)^2 \right) \Bigl\}.
	\end{align}
%
where 
\begin{align}
  I_{\rm B}(a_i^2) &= \int_0^\infty d x x^2\ln 
  \left[ 1- \exp \left( -\sqrt{x^2+a_i^2}\right) \right],
  \\
  I_{\rm F}(a_i^2) &= \int_0^\infty d x x^2\ln 
  \left[ 1+ \exp \left( -\sqrt{x^2+a_i^2}\right) \right],
\end{align}
%
with $a_i\equiv M_i(\varphi)/T$.
The SMEFT operators also contribute to the thermal effects, and we took into account such effects in our numerical calculations.
Under the high temperature approximation, these contributions are roughly given by
	\begin{align}
 &I_{\rm B}(a^2_i) \to -\frac{\pi^4}{45} + \frac{\pi^2}{12} a^2_i -  \frac{\pi}{6} (a^2_i)^{3/2} -  \frac{a^4_i}{32}\ln\left(\frac{a^2_i}{\alpha_B}\right)+\dots,\nonumber\\
 &I_{\rm F}(a^2_i)  \to \frac{7\pi^4}{360} - \frac{\pi^2}{24} a^2_i  -  \frac{a^4_i}{32}\ln\left(\frac{a^2_i}{\alpha_F}\right)+\dots, \label{eq:hitemp}
	\end{align}
%
where $\ln \alpha_B=2\ln4\pi -2\gamma_E+3/2$, $\ln\alpha_F=2\ln\pi -2\gamma_E+3/2$ and $\gamma_E$ is Euler constant.
To avoid the IR divergence from Matsubara zero-modes of bosons, we add ring diagram contributions~\cite{PhysRevD.45.2933} to the potential:
\begin{align}
V_T^{\rm ring}=\frac{T}{12\pi}\sum_{i= \varphi,W,Z}n_i \left( (M_i^2(\varphi, 0))^{3/2} - (M_i^2(\varphi, T))^{3/2}\right),
\end{align}
where $M_i^2(\varphi, T)=M_i^2(\varphi) + \Pi_i(T)$ and $\Pi_i(T)$ is the thermal self-energy defined as
\begin{align}
&\Pi_{\varphi}(T)\equiv  T^2\left(  \frac{ \lambda}{2} + \frac{ 3g^2}{16}  + \frac{{g'}^2}{16} +\frac{Y_t^2}{4} \right),
\\
&\Pi_{i}^{\rm (L,T)}(T)\equiv \frac{11T^2}{6}a_i^{\rm (L,T)}
\left(\begin{array}{cccccccc} 
 g^2  &0&0&0 \\
0& g^2  &0&0 \\
0&0& g^2 &0 \\
0&0&0& {g'}^2 
\end{array}\right),~~{\rm for}~i=W^1, W^2, W^3, B.
\end{align}
%
with $a^{\rm L}_i=1$, and $a^{\rm T}_i=0$.

The SFO-EWPT is quantified by the ratio $v_c/T_c$, where the $T_c$ and $v_c$ are determined by the two conditions
\begin{align}
    &\partial_{\varphi} V_{\rm eff}(\varphi,T_c)|_{\varphi=v_c}=0,\label{eq:cond1}
    \\
    &V_{\rm eff}(v_c,T_c)=V_{\rm eff}(0,T_c).\label{eq:cond2}
\end{align}
At the critical temperature $T_c$, two energetically degenerate minima arise at $\varphi=0$ and $\varphi=v_c$ and are separated by an energy barrier.
%
\begin{figure*}[t]
\centering
\includegraphics[width=0.47\textwidth]{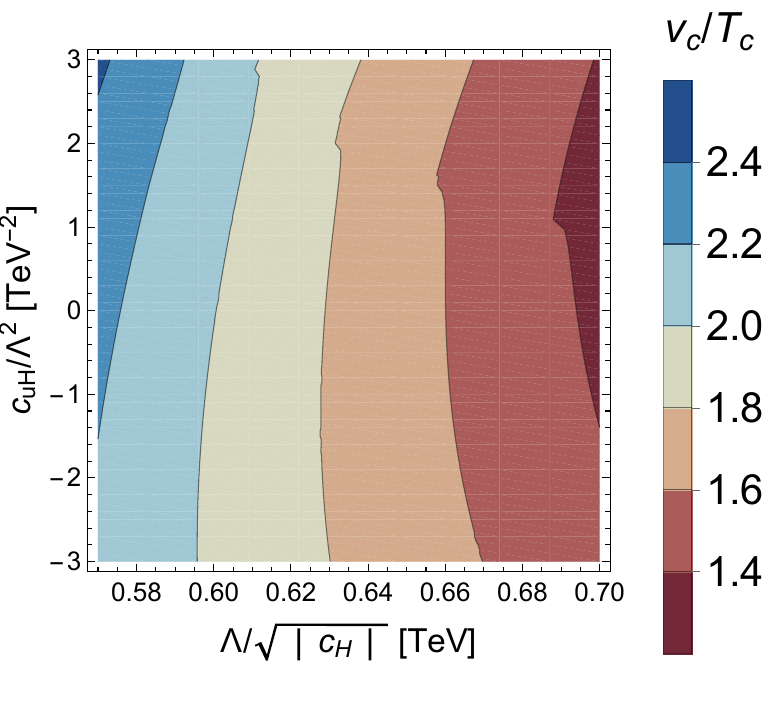}
\caption{
The ratio $v_c/T_c$ as a function of $\Lambda/\sqrt{|c_H|}$ and $c_{uH}/\Lambda^2$.
The other Wilson coefficients are taken to be zero, i.e., $c_{H\Box}=c_{HD}=0$.
The region of $v_c/T_c\geq 1$ satisfies the condition for the SFO-EWPT that can produce the GWs.
\label{fig:VCTC}
}
\end{figure*}
%
The impact of the SMEFT Wilson coefficients on the $v_c/T_c$ is estimated in Figs.~\ref{fig:VCTC}-\ref{fig:VCTCHBox} 
to clarify parameter spaces where the SFO-EWPT can arise. 
Here, we defined $C_H\equiv c_H/\Lambda^2$, $C_{uH}\equiv c_{uH}/\Lambda^2$, $C_{HD}\equiv c_{HD}/\Lambda^2$, and $C_{H\Box}\equiv c_{H\Box}/\Lambda^2$. 
In Figs.~\ref{fig:VCTC}-\ref{fig:VCTCHBox}, contours of $v_c/T_c$ are shown on the $\{\Lambda/\sqrt{|c_H|}$, $c_{uH}/\Lambda^2\}$,  $\{\Lambda/\sqrt{|c_H|}$, $c_{HD}/\Lambda^2\}$, and $\{\Lambda/\sqrt{|c_H|}$, $c_{H\Box}/\Lambda^2\}$ planes, respectively.
We assumed $c_{H\Box}=c_{HD}=0$, $c_{uH}=c_{H\Box}=0$ and $c_{uH}=c_{HD}=0$ for Fig.~\ref{fig:VCTC}-\ref{fig:VCTCHBox}, respectively.
From Figs.~\ref{fig:VCTC}-\ref{fig:VCTCHBox}, it is found that the $C_H$ tree level effect dominates the SFO-EWPT, and the other operators slightly change the Higgs potential.

For ease of understanding the numerical results, we qualitatively discuss the condition for achieving the SFO-EWPT. 
For simplicity of explanations, we omit $V_{\rm one\text{-}\rm loop}$, $V_T^{\rm ring}$, and second or higher order corrections for $C_{uH}$.  
From Eq.~\eqref{eq:HiggsP} and \eqref{eq:veffnum}, the effective potential at zero temperature is calculated as
\begin{align}
V_{\rm eff}(\varphi,0)=\frac{1}{2}a_2 \varphi^2 +\frac{1}{4}a_4 \varphi^4+\frac{1}{5}a_5 \varphi^5 +\frac{1}{6}a_6 \varphi^6,\label{eq:V0}
\end{align}
where
\begin{align}
    &a_2=\mu^2,~a_4=\lambda-\frac{4}{3}c_{\rm kin}^{(0)} \mu^2,
    ~a_5=-\frac{5}{4}c_{\rm kin}^{(1)}\mu^2,\notag
    \\
    &a_6=-\frac{3}{4}C_H-2 c_{\rm kin}^{(0)}\lambda-\frac{6}{5}c_{\rm kin}^{(2)}\mu^2 - \frac{9}{16\pi^2} C_{uH} Y^3_t \left(-1+\ln \frac{Y^2_t \varphi^2}{2v^2}\right).\label{eq:acoee}
\end{align}
The parameters $a_2,a_4$, i.e., $\mu^2$ and $\lambda$, are fixed by Eq.~\eqref{eq:Vpr} and \eqref{eq:Vprpr} as follows,
\begin{align}
    &\partial_{\varphi} V_{\rm eff}\left(\varphi, 0\right)|_{\varphi=v}=a_2 v +a_4 v^3+a_5 v^4 +a_6 v^5 =0,\label{eq:Vpr2}
    \\
    &\partial_{\varphi}^2 V_{\rm eff}\left(\varphi, 0\right)|_{\varphi=v}=a_2  +3 a_4 v^2+4a_5 v^3 +5 a_6 v^4 =m_{h}^2.\label{eq:Vprpr2}
\end{align}
By solving Eq.~\eqref{eq:Vpr2} and \eqref{eq:Vprpr2} with respect to the parameters $a_2$ and $a_4$, we obtain
\begin{align}
    V_{\rm eff}\left(\varphi, 0\right)=-\frac{1}{4}\left(m_{h}^2-a_5 v^3 -2 a_6 v^4\right)\varphi^2 +\frac{1}{4}\left(\frac{m_{h}^2}{2v^2}-\frac{3}{2}a_5 v-2 a_6 v^2\right)\varphi^4+\frac{1}{5}a_5 \varphi^5 +\frac{1}{6}a_6 \varphi^6,\label{eq:poSMEFT}
\end{align}
with $a_2=-(m^2_{h}-a_5 v^3-2 a_6 v^4)/2$, and $a_4=m^2_{h}/2v^2-3a_5 v/2 -2 a_6 v^2$.
For $a_5/2v+a_6\gtrsim m_{h}^2/2v^4$~\footnote{More precisely speaking, the quadratic and quartic terms change signs for $a_5/2v+a_6\gtrsim m_{h}^2/2v^4$ and $3a_5/2v+2a_6\gtrsim m_{h}^2/2v^4$, respectively.}, both quadratic and quartic terms of the potential at zero temperature change signs, and the SFO-EWPT can happen around the point:
\begin{align}
    \frac{a_5}{2v}+a_6\sim \frac{m_{h}^2}{2 v^4} \simeq (685~{\rm GeV})^{-2}.
\end{align}
Since the $\mathcal{O}_H$ operator dominates the SMEFT effect on the Higgs potential, $C_H$ takes a negative value for the SFO-EWPT.
Note here that the origin of the potential becomes a true minimum at zero temperature when $C_H$ is too large.
As studied in detail in Ref.~\cite{Ellis:2018mja}, the SFO-EWPT where percolation can be possible corresponds to $\Lambda/\sqrt{|c_H|}> 0.55$~TeV. 
%
\begin{figure*}[t]
\centering
\includegraphics[width=0.47\textwidth]{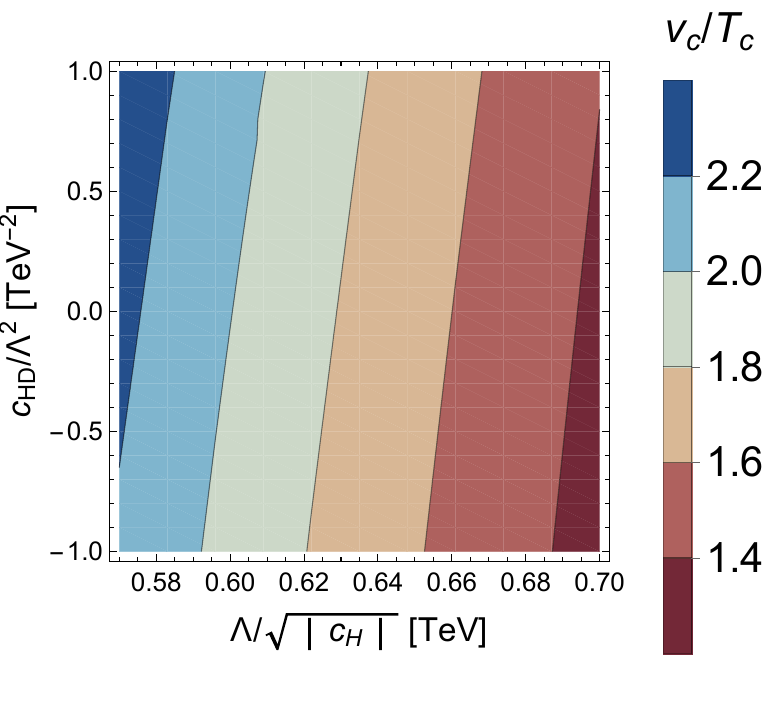}
\caption{
The same plot as Fig.~\ref{fig:VCTC} but as a function of $\Lambda/\sqrt{|c_H|}$ and $c_{HD}/\Lambda^2$.
$c_{uH}=c_{H\Box}=0$ is assumed.
\label{fig:VCTCHD}
}
\end{figure*}
%
Including the finite temperature effects, the effective Higgs potential in the high-temperature limit of Eq.~\eqref{eq:hitemp} is approximately given by
\begin{align}
    V_{\rm eff}(\varphi,T)\sim\frac{1}{2}A_2\varphi^2-\frac{1}{2\sqrt{2}} E T \varphi^3 +\frac{1}{4} A_4 \varphi^4,
\end{align}
where $A_2$ is a coefficient of the quadratic term involving the thermal corrections, $E$ denotes a coefficient of the cubic term from the thermal effects of the SM bosons, and $A_4\equiv a_4 +4a_5\varphi/5+2 a_6 \varphi^2/3$.
From Eq.~\eqref{eq:cond1} and \eqref{eq:cond2}, the $T_c$ and $v_c$ are calculated as follows,
\begin{align}
    \frac{v_c}{T_c} = \frac{E}{\sqrt{2}A_4}\sim \frac{E}{\sqrt{2}\left(m^2_{h}/(2v^2)-3a_5 v/2-2 a_6 v^2 \right)}.\label{eq:vcTc}
\end{align}
Therefore, the SFO-EWPT requires ${v_c}/{T_c}\geq 1$, i.e., $a_5/2v+a_6\gtrsim m_{h}^2/2v^4$. 
Since $a_5$ is generated by top-quark loop effects and is a sub-dominant effect, from Eq.~\eqref{eq:vcTc}, it is clear that the parameter $a_6$, i.e., $C_H$, dominates the SMEFT effects on $v_c/T_c$, as shown in Figs.~\ref{fig:VCTC}-\ref{fig:VCTCHBox}.
%

\begin{figure*}[t]
\centering
\includegraphics[width=0.47\textwidth]{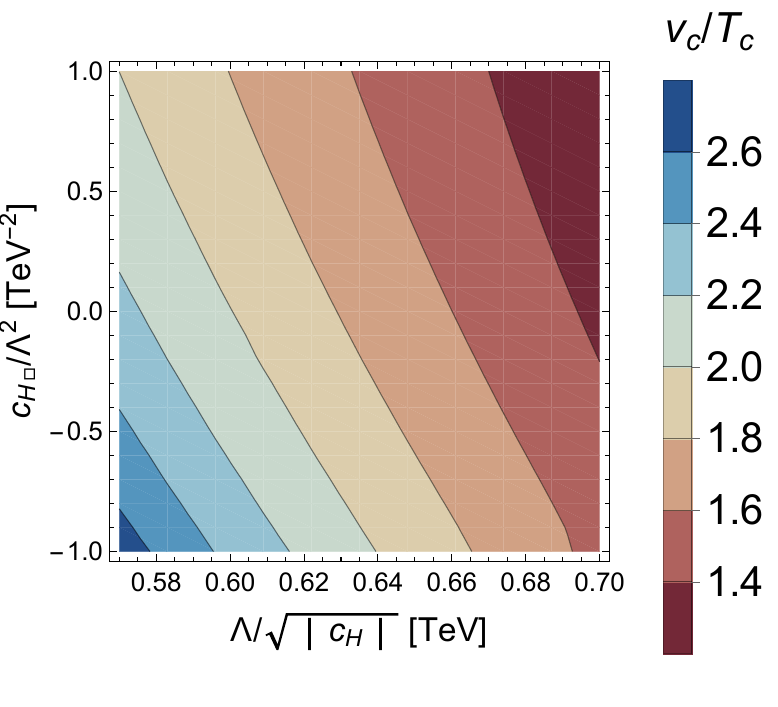}
\caption{
The same plot as Fig.~\ref{fig:VCTC} but as a function of $\Lambda/\sqrt{|c_H|}$ and $c_{H\Box}/\Lambda^2$.
$c_{uH}=c_{HD}=0$ is assumed.
\label{fig:VCTCHBox}
}
\end{figure*}

\section{GW spectrum and statistical analysis}
\label{sec:GWsp}

The GWs can arise from the SFO-EWPT, and their spectrum can be determined by a few quantities, such as the latent heat and the bubble nucleation rate.
As explained later, the produced GW spectrum is characterized by four parameters, and the Wilson coefficients of the SMEFT are encoded into the four parameters.
Future GW experiments such as LISA, DECIGO, and BBO are potentially sensitive to the GW spectrum from the SFO-EWPT, and the Wilson coefficients of the SMEFT may be constrained. 
In this section, we review Refs.~\cite{Caprini:2019egz,Hindmarsh:2017gnf,PhysRevD.101.089902,Espinosa:2010hh,Seto:2005qy,Hashino:2018wee,Yagi:2011wg,Klein:2015hvg} and summarize the GW spectrum from the SFO-EWPT and analysis to derive the constraints on the Wilson coefficients.

\subsection{GW spectrum from first-order phase transition}
The GW from the first-order phase transition is mainly characterized by four parameters: $T_t$, $\alpha$, $\beta/H$, and $v_b$.
We refer to these parameters as the phase transition parameters.
$T_t$ is the temperature during bubble percolation for the phase transition to complete and is defined by
	\begin{align} 
	\Gamma/H^4|_{T=T_t}=1,\label{eq:GamH}
	\end{align}
where $H=8\pi^3g_* T^4/90m_{\rm Pl}^2$ is the Hubble parameter in a radiation-dominated epoch with the Plank mass $m_{\rm Pl}$ and degrees of freedom in the plasma $g_*=106.75$, and $\Gamma$ is a bubble nucleation rate per unit volume and unit time:
	\begin{align} 
	\Gamma\simeq T^4\left(\frac{S_3}{2\pi T} \right)^{3/2} \exp (-S_3/T),\label{eq:Gam}
	\end{align}
with a 3-dimensional Euclidean action $S_3$, i.e., $O(3)$ symmetric bounce solution.
In the numerical calculation, the bounce solution is calculated by the {\tt AnyBubble}~\cite{Masoumi:2016wot} package.
Combining Eq.~\eqref{eq:GamH} and \eqref{eq:Gam}, $T_t$ is numerically calculated.
The second parameter $\alpha$ is a ratio of the released latent heat $\epsilon$ and the background plasma energy density $\rho_{\rm rad}(T)=(\pi^2/30)g_* T^4$ at $T=T_t$ as follows,  
	\begin{align} 
	\alpha\equiv \epsilon(T_t)/ \rho_{\rm rad}(T_t).
	\end{align}
Here, the released latent heat is defined as 
 \begin{align}
 \label{latenth}
  \epsilon(T)
  =  \Delta V_{\rm eff} -T
  \frac{\partial  \Delta V_{\rm eff} }{\partial T},\quad  \Delta V_{\rm eff} =  V_{\rm eff}(\varphi_-(T),T) - V_{\rm eff}(\varphi_+(T),T),
\end{align}
where $V_{\rm eff}$ is defined in Eq.~\eqref{eq:veffnum}, and $\varphi_{+}$ and $\varphi_{-}$ denote the order parameters for the broken and unbroken phases, respectively.
The third parameter $\beta/H$ represents the inverse of the duration of the phase transition and is defined as
	\begin{align} 
	\frac{\beta}{H}\equiv T_t\left.\frac{d}{dT}\left(\frac{S_3}{T}\right)\right|_{T=T_t}.
	\end{align}
The last parameter $v_b$ is the bubble wall velocity, which is the speed of the bubble wall in the rest frame of the plasma far from the wall.
In the following numerical analysis, we choose a benchmark point of $v_b=0.3$, where the electroweak baryogenesis involving $C_{uH}$ is also possible~\cite{Bodeker:2004ws}.
In Appendix~\ref{app:vel}, a different choice of the benchmark point of $v_b$ is considered.

 The GW from the first-order phase transition arises from three sources: bubble collision, plasma turbulence, and compression wave of plasma.
In particular, the compression wave of plasma is the dominant source of the GW spectrum, so we focus only on it in the following numerical calculations.  
The fitting function for the numerical simulations of the GW spectrum generated by a phase transition during the radiation era is expressed as~\cite{Caprini:2019egz,Hindmarsh:2017gnf,PhysRevD.101.089902}
%
	\begin{align}
  \Omega_{\rm comp} (f) 
  &= 2.061 F_{\rm gw,0}\tilde{\Omega}_{\rm gw} \left(\frac{f}{\tilde{f}_{\rm comp}}\right)^3
  \left(\frac{7}{4+3(f/\tilde{f}_{\rm comp})^2}\right)^{7/2}\notag
  \\
  &\times\begin{cases}
  \left(\frac{\kappa_v\alpha}{1+\alpha}\right)^2  (H_*R_*)
   ,~~~{\rm for}~H_*R_* \leq \sqrt{\frac{3}{4}\kappa_v\alpha/(1+\alpha)}
   \\
    \left(\frac{\kappa_v\alpha}{1+\alpha}\right)^{3/2} (H(T_t)R_*)^2,~~~{\rm for}~ \sqrt{\frac{3}{4}\kappa_v\alpha/(1+\alpha)} < H_*R_* 
  \end{cases}  \label{eq:GWspcom} 
	  \end{align}
where an efficiency factor $\kappa_v$ ~\cite{Espinosa:2010hh} is given by a function of $\alpha$ and $v_b$ as follows,
\begin{align}
  \kappa_v(v_b, \alpha)\simeq
  \begin{cases}
    & \frac{ c_s^{11/5}\kappa_A \kappa_B }{(c_s^{11/5} -  v_b^{11/5} )\kappa_B
      +  v_b c_s^{6/5} \kappa_A}\quad \text{for}~ v_b \lesssim c_s  \\
    & \kappa_B + ( v_b - c_s) \delta\kappa 
    + \frac{( v_b - c_s)^3}{ (v_J - c_s)^3} [ \kappa_C - \kappa_B -(v_J - c_s) 
    \delta\kappa ]
    \quad \text{for}~ c_s <  v_b < v_J \\
    & \frac{ (v_J - 1)^3 v_J^{5/2}  v_b^{-5/2}
      \kappa_C \kappa_D }
    {[( v_J -1)^3 - ( v_b-1)^3] v_J^{5/2} \kappa_C
      + ( v_b - 1)^3 \kappa_D }
    \quad \text{for}~ v_J \lesssim v_b 
  \end{cases},    
\end{align}
%
where 
\begin{align}
  \kappa_A 
  &\simeq v_b^{6/5} \frac{6.9 \alpha}{1.36 - 0.037 \sqrt{\alpha} + \alpha},
  \quad
  \kappa_B 
  \simeq \frac{\alpha^{2/5}}{0.017+ (0.997 + \alpha)^{2/5} },
    \nonumber \\
  \kappa_C 
  &\simeq \frac{\sqrt{\alpha}}{0.135 + \sqrt{0.98 + \alpha}},
  \quad
  \kappa_D 
  \simeq \frac{\alpha}{0.73 + 0.083 \sqrt{\alpha} + \alpha}.
\end{align}
%
Here, $c_s$ is the velocity of sound ($c_s=0.577$), and 
\begin{align}
    v_J=\frac{\sqrt{2/3\alpha +\alpha^2}+\sqrt{1/3}}{1+\alpha},~~~    \delta \kappa\simeq -0.9 \ln \frac{\sqrt{\alpha}}{1+\sqrt{\alpha}}.
\end{align}
Also, $F_{\rm gw,0} = 3.57 \times 10^{-5} \left(100/g_\ast\right)^{1/3}$, $\tilde{\Omega}_{\rm gw}=1.2 \times 10^{-2}$, $H_*R_*=(8\pi)^{1/3}(\beta/H)^{-1} $ max($c_s,v_b$) and $\tilde{f}_{\rm comp}$ is the peak frequency given by
	\begin{align}
  \tilde{f}_{\rm comp} \simeq 26 \left(\frac{1}{H_*R_*}\right) 
  \left(\frac{T_t}{100~{\rm GeV}}\right)
  \left(\frac{g_\ast}{100}\right)^{1/6} ~\mu{\rm Hz}.
	\end{align}
%
As explained in the previous sections, the Wilson coefficients of the SMEFT contribute to the effective Higgs potential, and the SMEFT effects are encoded into the phase transition parameters.
The point is that the GW spectrum from the first-order phase transition is determined from the Wilson coefficients through the phase transition parameters.

\begin{figure*}[t]
\centering
\includegraphics[width=0.8\textwidth]{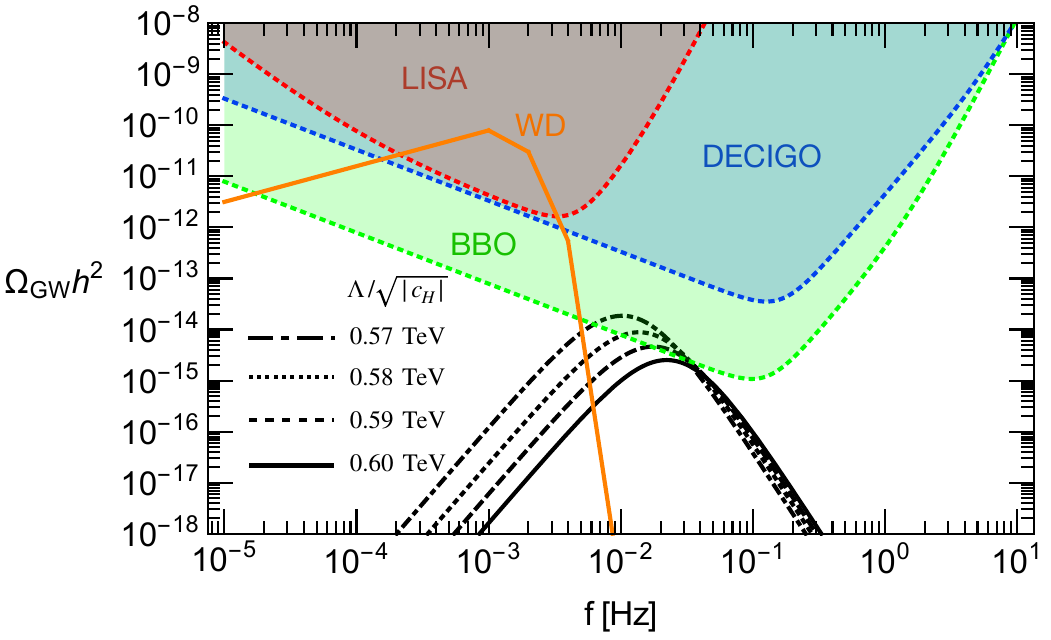}
\caption{
The GW spectra from the SFO-EWPT by the SMEFT $\mathcal{O}_H$ operator effects.
The black curves correspond to $\Lambda/\sqrt{|c_H|}=0.57$ TeV (dot-dashed), $0.58$ TeV (dotted), $0.59$ TeV (dashed), and $0.60$ TeV (solid), respectively, assuming $c_{HD}=c_{H\Box}=c_{uH}=0$.
The colored regions represent the sensitivity regions of LISA (red), DECIGO (blue), and BBO (green).
The orange curve denotes a foreground coming from compact white dwarf binaries in our Galaxy.
\label{fig:GWsp}
}
\end{figure*}\subsection{Statistical analysis in GW experiments}
The Wilson coefficients of the SMEFT are potentially measured by future GW observations.
The error of the Wilson coefficients denotes room for the NP effects when its central value is zero consistently within the error.
We evaluate the confidence interval of the Wilson coefficients in the GW observation by the Fisher matrix analysis and investigate the sensitivity of future GW experiments to the SMEFT operators.
The sensitivity of future GW experiments to the Wilson coefficient is well characterized by the signal-to-noise ratio (SNR). 
The SNR for the observation of GW spectrum is obtained as~\cite{Seto:2005qy,Hashino:2018wee}
\begin{align}\label{eq:SNR}
\mathrm{SNR}= \sqrt{\delta\times T_{\rm obs} \int^{\infty}_{0}df \left[\frac{\Omega_{\rm GW}(f)}{\Omega_{\rm sen}(f)}\right]^2},
\end{align}
where $\Omega_{\rm GW}\simeq \Omega_{\rm comp}$, $T_{\rm obs}$ is the observation period, $\Omega_{\rm sen}\equiv (2\pi^2 f^3/3H_0^2)\cdot S_{\rm eff}(f)$ is the sensitivity of experiments, which are summarized later, and $\delta$ is the number of independent channels for the experiments, i.e., $\delta=2$ for cross-correlated detectors such as DECIGO and BBO, and $\delta=1$ for LISA.
The logarithm of the likelihood function is approximated to~\cite{Seto:2005qy,Hashino:2018wee}
%
\begin{align}
  \delta \chi^2(\{p\},\{\hat{p}\})\simeq {\cal F}_{ab}(p_a-\hat{p_a})(p_b-\hat{p_b}),\label{eq:chi}
\end{align}
where the parameter set $\{p\}$ denotes the Wilson coefficients of the SMEFT, $\{\hat{p}\}$ is the set of fiducial values of $\{p\}$, and ${\cal F}_{ab}$ represents the Fisher information matrix~\cite{Seto:2005qy} defined as follows,
\begin{align}
{\mathcal F}_{ab}
&=
2T_{\rm obs}
\int_0^\infty df
~
\frac{\partial_{p_a} S_h(f,\left\{ \hat{p} \right\}) \partial_{p_b} S_h(f,\left\{ \hat{p} \right\})}
{\left[ S_{\rm eff}(f) + S_h(f,\left\{ \hat{p} \right\}) \right]^2},
\label{eq:FabSeff}
\end{align}
with the power spectrum
\begin{align}
S_h(f)
&= 
\frac{3H_0^2}{2\pi^2}
\frac{1}{f^3}
\Omega_{\rm GW}(f).\label{eq:pow}
\end{align}
Throughout this work, we evaluate the confidence intervals on two-dimensional planes and consider three parameter sets:  $\{C_H, C_{uH}\}$, $\{C_H, C_{HD}\}$, and $\{C_H, C_{H\Box}\}$. 
The 95\% C.L. interval of the Wilson coefficients denotes a contour of $\delta \chi^2 = 6.0$ in the two-dimensional plane. 
The effective sensitivities of each experiment are evaluated as~\cite{Yagi:2011wg,Klein:2015hvg} 
\begin{itemize}
\item LISA

\begin{align}
    S_{\rm eff}(f)=\frac{20}{3}\frac{4 S_{\rm acc}(f)+S_{\rm sn}(f)+S_{\rm omn}(f)}{L^2}\left[1+\left(\frac{f}{0.41 c/2L}\right)^2\right],
\end{align}
with $L=5\times 10^9$~m and,
\begin{align}
    &S_{\rm acc}(f)=9\times 10^{-30} \frac{1}{(2\pi f/1{\rm Hz})^4}\left(1+\frac{10^{-4}}{f/1{\rm Hz}}\right)~{\rm m^2 Hz^{-1}},
    \\
    &S_{\rm sn}(f)=2.96\times 10^{-23}~{\rm m^2 Hz^{-1}},
    \\
    &S_{\rm omn}(f)=2.65\times 10^{-23}~{\rm m^2 Hz^{-1}}.
\end{align}

\item
DECIGO
\begin{align}
S_{\rm eff}(f) 
&=
\left[
\frac{}{}
7.05 \times 10^{-48} 
\left[
1 + (f / f_p)^2
\right]
\right.
\nonumber \\
&~~~~
\left.
+ 
4.8 \times 10^{-51} 
\frac{(f / 1{\rm Hz})^{-4}}{1 + (f / f_p)^2} 
+
5.33 \times 10^{-52}
(f / 1{\rm Hz})^{-4}
\right]
~{\rm Hz^{-1}},
\label{eq:SeffDECIGO}
\end{align}
with $f_p = 7.36$~Hz.
\item
BBO 
\begin{align}
S_{\rm eff}(f)
&=
\left[
2.00 \times 10^{-49} 
(f / 1{\rm Hz})^2
+ 
4.58 \times 10^{-49}
+
1.26 \times 10^{-52}
(f / 1{\rm Hz})^{-4}
\right]
~{\rm Hz^{-1}}.
\label{eq:SeffBBO}
\end{align}
\end{itemize}

It is known that the stochastic GWs from astrophysical sources can be a foreground.
In our numerical calculations, we add a foreground from compact white dwarf binaries in our Galaxy in the milli-Hertz regime to the effective sensitivity of each experiment.
The noise spectrum of the white dwarf is evaluated as~\cite{Klein:2015hvg}
\begin{align}
S'_{\rm WD}(f)=
\begin{cases}
(20/3) (f / 1~{\rm Hz})^{-2.3} \times 10^{-44.62}~{\rm Hz^{-1}}&\equiv S^{(1)}_{\rm WD}(f)
\quad(10^{-5}~{\rm Hz} < f < 10^{-3}~{\rm Hz}), \\
(20/3) (f / 1~{\rm Hz})^{-4.4} \times 10^{-50.92}~{\rm Hz^{-1}}&\equiv S^{(2)}_{\rm WD}(f)
\quad (10^{-3}~{\rm Hz} < f < 10^{-2.7}~{\rm Hz}), \\
(20/3) (f / 1~{\rm Hz})^{-8.8} \times 10^{-62.8}~{\rm Hz^{-1}}&\equiv S^{(3)}_{\rm WD}(f)
\quad (10^{-2.7}~{\rm Hz} < f < 10^{-2.4}~{\rm Hz}), \\
(20/3) (f / 1~{\rm Hz})^{-20.0} \times 10^{-89.68}~{\rm Hz^{-1}}&\equiv S^{(4)}_{\rm WD}(f)
\quad(10^{-2.4}~{\rm Hz} < f < 10^{-2}~{\rm Hz}). 
 \end{cases}
\end{align}
In the numerical calculations, we adopt the following noise spectrum
\begin{align}
    S_{\rm WD}(f)=\frac{1}{1/S_{\rm WD}^{(1)}(f)+1/S_{\rm WD}^{(2)}(f)+1/S_{\rm WD}^{(3)}(f)+1/S_{\rm WD}^{(4)}(f)}.
\end{align}
In Fig.~\ref{fig:GWsp}, the sensitivity regions for LISA, DECIGO, and BBO are shown as red-shaded, blue-shaded, and green-shaded regions, respectively.
The orange curve denotes the noise spectrum of the white dwarf.

\begin{figure*}[t]
\centering
\includegraphics[width=0.47\textwidth]{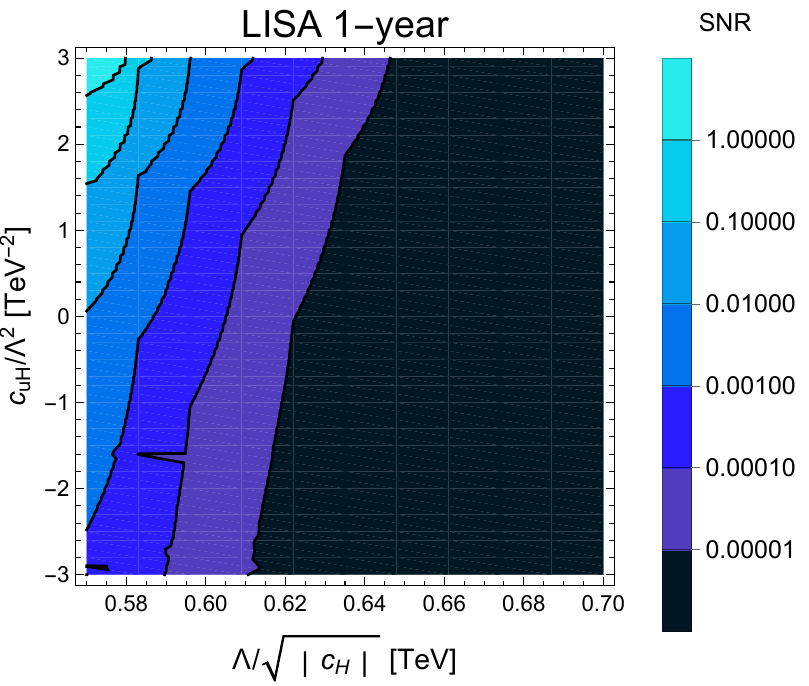}
\includegraphics[width=0.45\textwidth]{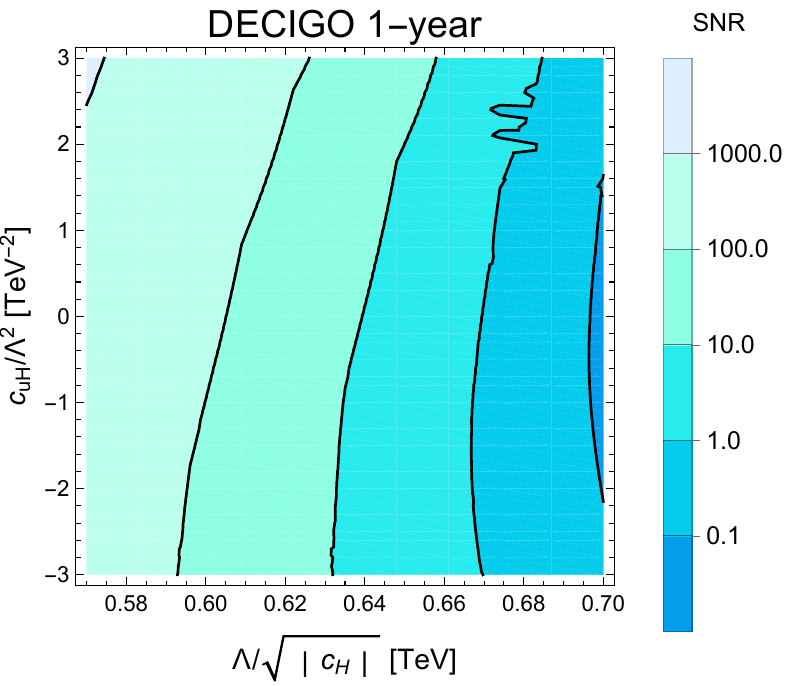}
\includegraphics[width=0.47\textwidth]{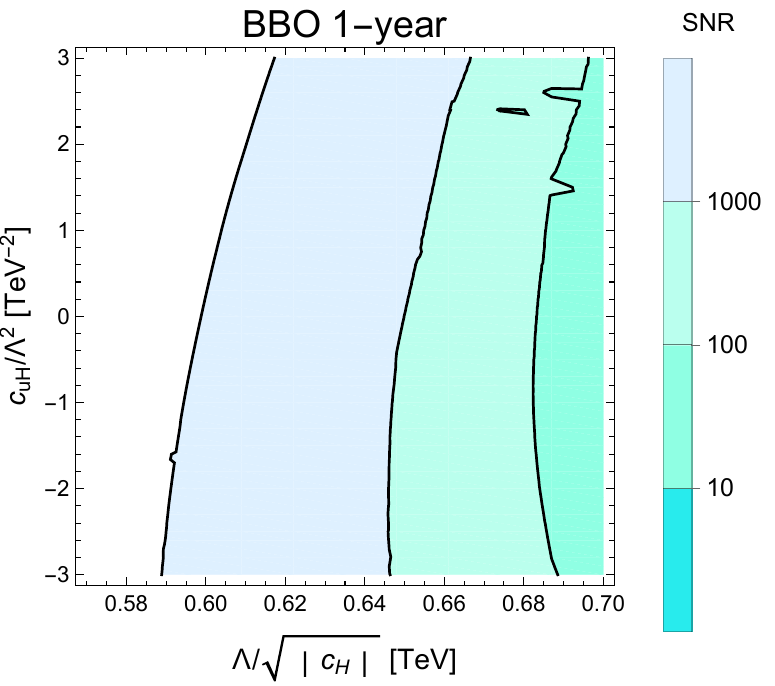}
\caption{
The SNR of the future GW experiments on the $(\Lambda/\sqrt{|c_H|}, c_{uH}/\Lambda^2)$ plane, assuming $v_b=0.3$, $c_{H\Box}=c_{HD}=0$, and 1-year statistics at LISA (upper left), DECIGO (upper right), and BBO (bottom).
The solid curves denote the contours of SNR.
\label{fig:SNRCuH}
}
\end{figure*}

\begin{figure*}[t]
\centering
\includegraphics[width=0.47\textwidth]{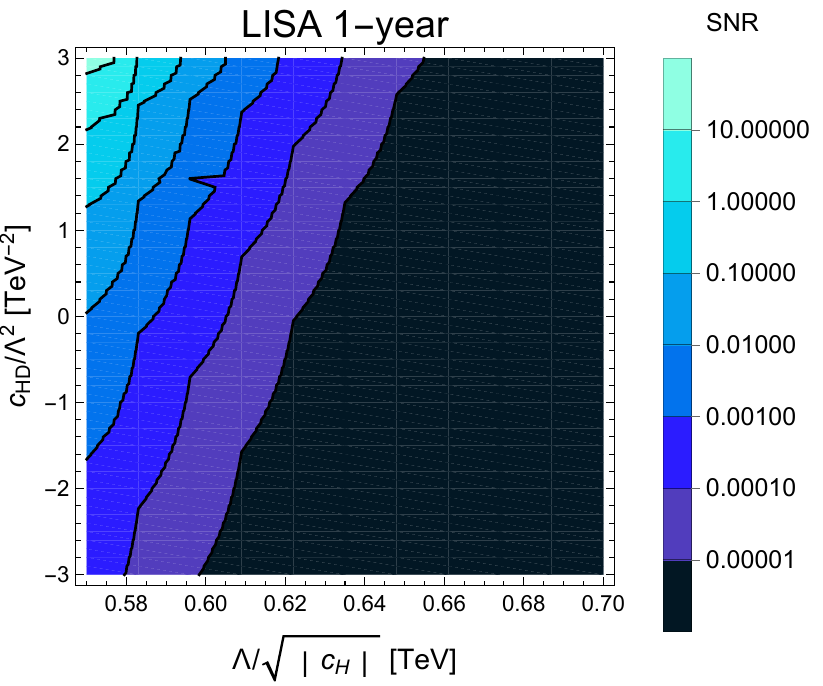}
\includegraphics[width=0.45\textwidth]{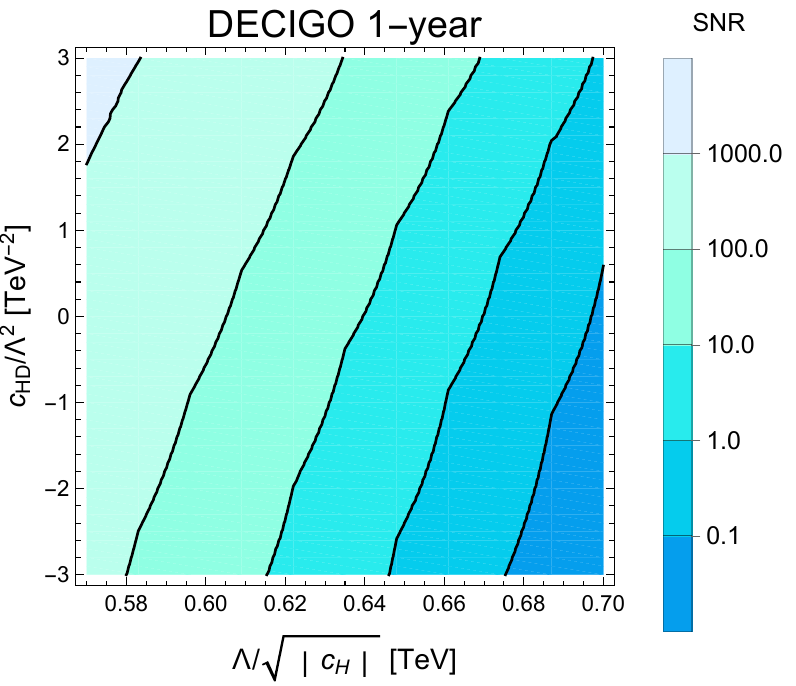}
\includegraphics[width=0.47\textwidth]{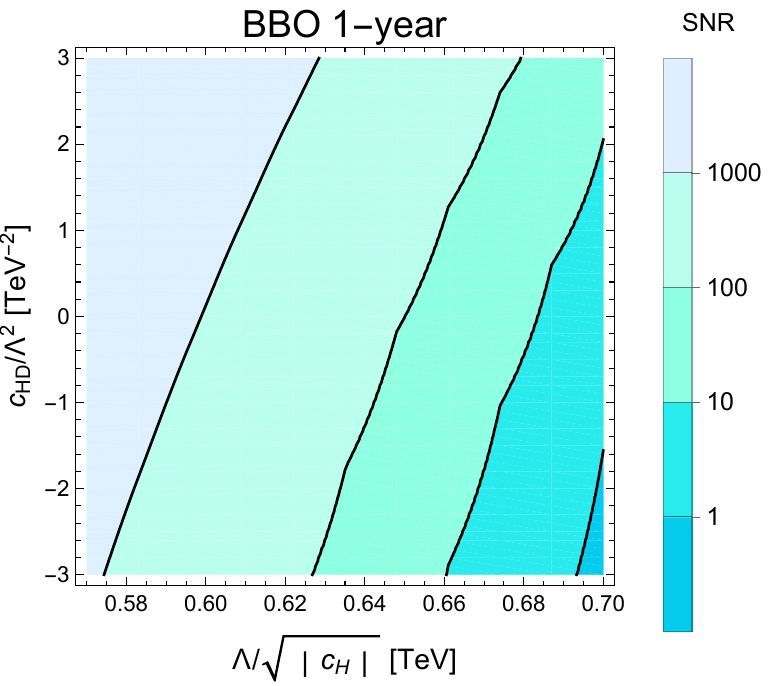}
\caption{
The same plot as Fig.~\ref{fig:SNRCuH} but on the $(\Lambda/\sqrt{|c_H|}, c_{HD}/\Lambda^2)$ plane, assuming $c_{uH}=c_{H\Box}=0$.
\label{fig:SNRCHD}
}
\end{figure*}

\section{Numerical results}
\label{sec:res}

Figs.~\ref{fig:VCTC}-\ref{fig:VCTCHBox} show numerical results of ratio $ v_c/T_c$ as functions of $\{\Lambda /\sqrt{|c_H|}$, $c_{uH}/\Lambda^2\}$, $\{\Lambda /\sqrt{|c_H|}$, $c_{HD}/\Lambda^2\} $, and $\{\Lambda /\sqrt{|c_H|}$, $c_{H\Box}/\Lambda^2\}$, respectively.
We assumed $c_{H\Box}=c_{HD}=0$, $c_{uH}=c_{H\Box}=0$, and $c_{uH}=c_{HD}=0$ for Figs.~\ref{fig:VCTC}-\ref{fig:VCTCHBox}, respectively.
For all figures, $v_c/T_c$ is larger than one in the regions $\Lambda/\sqrt{|c_H|}\lesssim 0.7$ TeV, and the GWs can arise from the SFO-EWPT.
Fig.~\ref{fig:GWsp} shows the GW spectrum from the SFO-EWPT achieved by the SMEFT operator $\mathcal{O}_H$.
The $\Lambda/\sqrt{|c_H|}$ dependence of the GW spectrum are shown, and the four black curves correspond to $\Lambda/\sqrt{|c_H|}=0.57$ TeV (dot-dashed), $0.58$ TeV (dotted), $0.59$ TeV (dashed), and $0.60$ TeV (solid). 
The colored shaded regions represent the effective sensitivities $(2\pi^2 f^3/ 3 H_0^2)\cdot S_{\rm eff}(f)$ of LISA (red), DECIGO (blue), and BBO (green). 
The orange curve corresponds to the foreground from compact white dwarf binaries in our Galaxy in the milli-Hertz regime.

\begin{figure*}[t]
\centering
\includegraphics[width=0.47\textwidth]{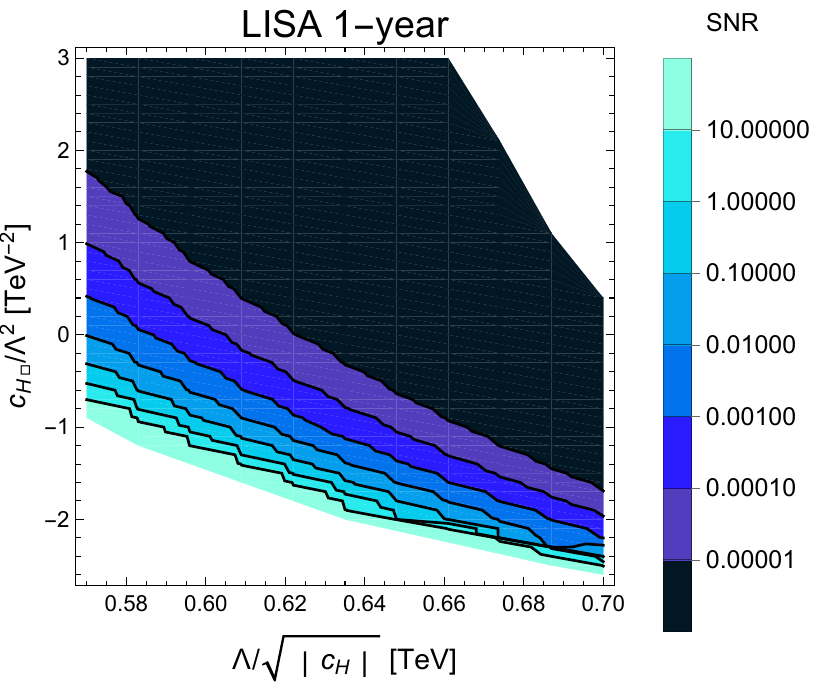}
\includegraphics[width=0.45\textwidth]{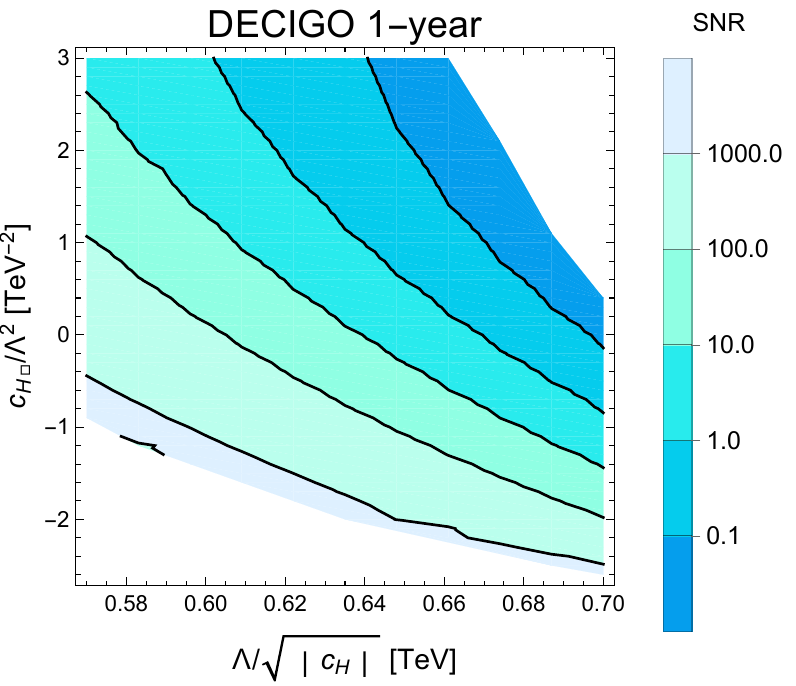}
\includegraphics[width=0.47\textwidth]{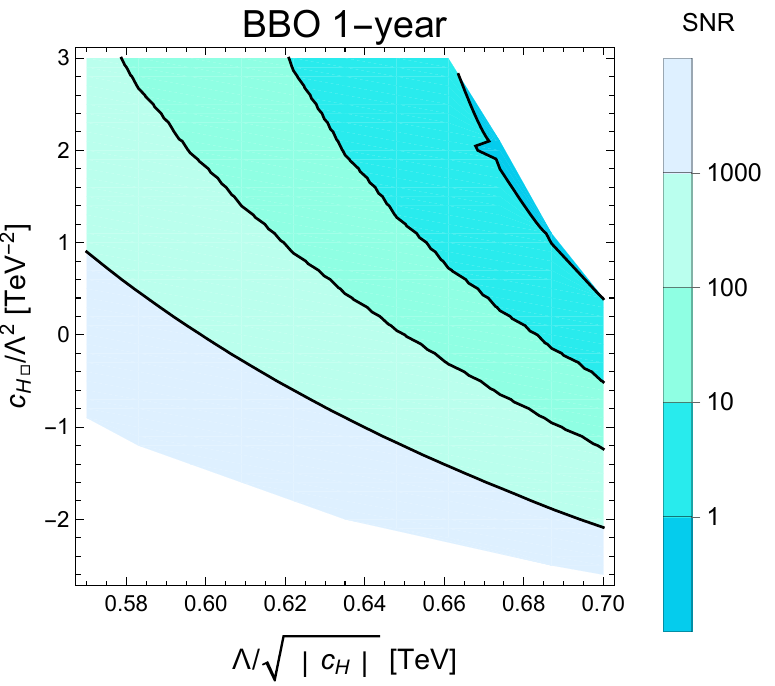}
\caption{
The same plot as Fig.~\ref{fig:SNRCuH} but on the $(\Lambda/\sqrt{|c_H|}, c_{H\Box}/\Lambda^2)$ plane, assuming $c_{uH}=c_{HD}=0$.
\label{fig:SNRCHBox}
}
\end{figure*}

We numerically evaluated the three SMEFT operators $\mathcal{O}_{uH}$, $\mathcal{O}_{H\Box}$, and $\mathcal{O}_{HD}$ effects on the GWs produced by $\mathcal{O}_{H}$.
To quantitatively evaluate the sensitivity of the future GW experiments to the three SMEFT operator effects, we calculated the SNRs of LISA, DECIGO, and BBO with $T_{\rm obs} =$ 1-year statistics and $v_b=0.3$, as shown in Figs.~\ref{fig:SNRCuH}-\ref{fig:SNRCHBox} on the $\{\Lambda/\sqrt{|c_H|}, c_{uH}/\Lambda^2\}$, $\{\Lambda/\sqrt{|c_H|}, c_{HD}/\Lambda^2\}$, and $\{\Lambda/\sqrt{|c_H|}, c_{H\Box}/\Lambda^2\}$ planes, respectively.
We assumed $c_{H\Box}=c_{HD}=0$, $c_{uH}=c_{H\Box}=0$, and $c_{uH}=c_{HD}=0$ for Figs.~\ref{fig:SNRCuH}-\ref{fig:SNRCHBox}, respectively.
In all figures of the LISA experiment, the SNRs for $c_{uH}=c_{H\Box}=c_{HD}=0$ are smaller than $\sim 10$, which corresponds to a typical value for the precisely measurable $c_{uH}, c_{H\Box}$, and $c_{HD}$ as explained later.
In the DECIGO experiment under the condition $c_{uH}=c_{H\Box}=c_{HD}=0$, the SNR of Figs.~\ref{fig:SNRCuH}-\ref{fig:SNRCHBox} is larger than $\sim 10$ for $\Lambda/\sqrt{|c_H|}\lesssim 0.64$ TeV.
The SNR of BBO is $\sim 10$ times larger than that of DECIGO; see Fig.~\ref{fig:GWsp}.
In Fig.~\ref{fig:SNRCHBox}, the measurable GWs are not generated in the white-shaded regions.
This is because, in the small $\Lambda/\sqrt{|c_H|}$ and $c_{H\Box}/\Lambda^2$ region, $a_6$ is too large to yield the SFO-EWPT, and in the large $\Lambda/\sqrt{|c_H|}$ and $c_{H\Box}/\Lambda^2$ region, $a_6$ is too small to yield the SFO-EWPT involving the measurable GWs; see Fig.~\ref{fig:VCTCHBox}. 
Comparing Figs.~\ref{fig:VCTC}-\ref{fig:VCTCHBox} with Figs.~\ref{fig:SNRCuH}-\ref{fig:SNRCHBox}, it is found that the behaviors of the contours of SNR are similar to those of $ v_c/T_c$.
This is because the phase transition parameters highly depend on $v_c/T_c$, e.g., $\alpha\propto (v_c/T_c)^2$~\cite{Espinosa:2010hh}. 

\begin{figure*}[t]
\centering
\includegraphics[width=0.45\textwidth]{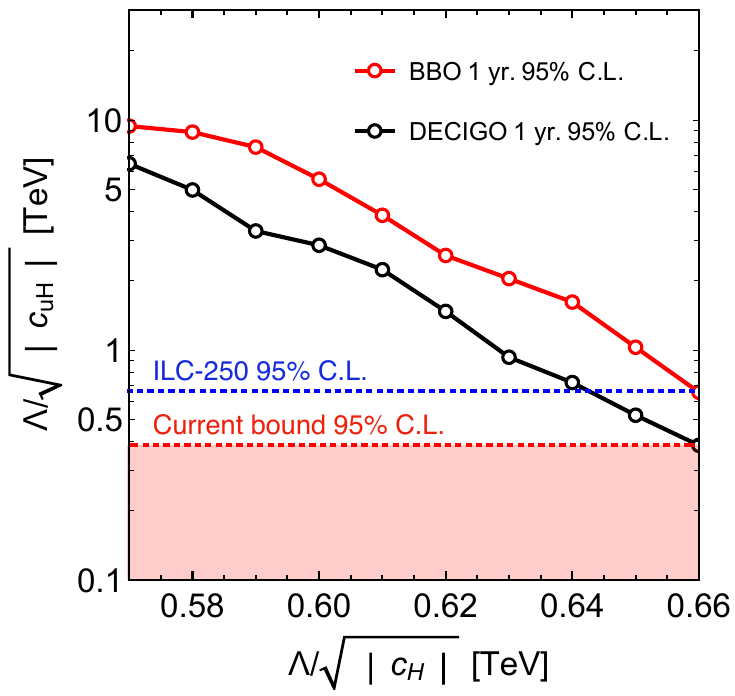}
\includegraphics[width=0.46\textwidth]{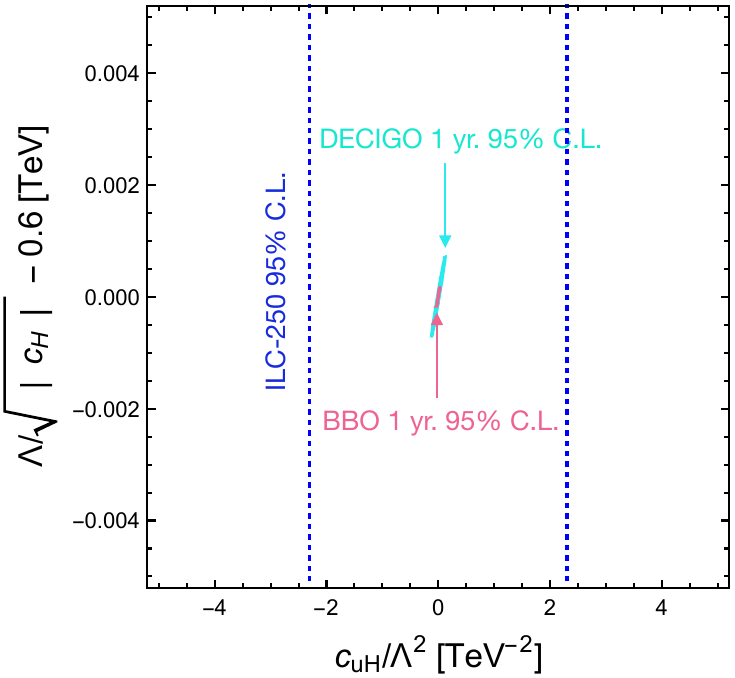}
\caption{
Left panel: Sensitivity reach of the DECIGO (black) and BBO (red) experiments to $\Lambda/\sqrt{|c_{uH}|}$ as a function of the central value of $\Lambda/\sqrt{|c_H|}$, assuming $v_b=0.3$, the central value of $c_{uH}/\Lambda^2=0$, and $T_{\rm obs} = $ 1-year statistics.
The vertical axis of the left panel is not the central value of $\Lambda/\sqrt{|c_{uH}|}$, but the magnitude of 95\% confidence intervals of $\Lambda/\sqrt{|c_{uH}|}$.
The red-dotted line is the current bound at 95\% C.L.~\cite{Ethier:2021bye}.
The dashed blue line shows a sensitivity reach of the ILC-250 at 95\% C.L.~\cite{DeBlas:2019qco,deBlas:2022ofj}.
Right panel: 95\% C.L. confidence regions for DECIGO (blue-shaded) and BBO (red-shaded) with 1-year statistics, assuming the central values of $c_{uH}/\Lambda^2=0$ and $\Lambda/\sqrt{|c_H|}=0.60~{\rm TeV}$.
The confidence interval for $c_{uH}/\Lambda^2$ direction in the right panel corresponds to one of the points in the left panel.
\label{fig:sens1}
}
\end{figure*}

We evaluate the confidence intervals in future GW experiments for the SMEFT Wilson coefficients by performing the Fisher matrix analysis.
Figs.~\ref{fig:sens1}-\ref{fig:sens3} show the 95\% confidence intervals, i.e., $\delta \chi^2=6.0$ in Eq.~\eqref{eq:chi}, in the DECIGO and BBO experiments with $T_{\rm obs}=$ 1-year statistics and $v_b=0.3$ for the SMEFT Wilson coefficients $C_{uH}$, $C_{H\Box}$, $C_{HD}$, and $C_H$.
In the right panels of Figs.~\ref{fig:sens1}-\ref{fig:sens3}, the 95\% confidence intervals are shown on the $\{c_{uH}/\Lambda^2,$ $\Lambda/\sqrt{|c_H|}\}$, $\{c_{HD}/\Lambda^2,$ $\Lambda/\sqrt{|c_H|}\}$, and $\{c_{H\Box}/\Lambda^2,$ $\Lambda/\sqrt{|c_H|}\}$ planes, respectively.
We assumed the central values in the right panels of Figs.~\ref{fig:sens1}-\ref{fig:sens3} as $\{c_{uH}/\Lambda^2=0, \Lambda/\sqrt{|c_H|}=0.60~{\rm TeV}\}$, $\{c_{HD}/\Lambda^2=0, \Lambda/\sqrt{|c_H|}=0.60~{\rm TeV}\}$, and $\{c_{H\Box}/\Lambda^2=0, \Lambda/\sqrt{|c_H|}=0.60~{\rm TeV}\}$, respectively.
The shaded blue and red regions denote the 95\% confidence regions in the DECIGO and BBO experiments, respectively.
The confidence intervals for $c_{uH}/\Lambda^2$, $c_{HD}/\Lambda^2$, and $c_{H\Box}/\Lambda^2$ directions represent allowed NP effects because their central values are assumed to be zero.
The dotted blue lines denote the sensitivity reach of the ILC-250 at 95\% C.L.~\cite{DeBlas:2019qco,deBlas:2022ofj}.
From Figs.~\ref{fig:sens1}-\ref{fig:sens3}, it is found that the sensitivity of the future GW observations may be higher than that of the future collider experiment.  
The left panels of Figs.~\ref{fig:sens1}-\ref{fig:sens3} show the sensitivity reach of the DECIGO and BBO experiments to $\Lambda/\sqrt{|c_{uH}|}$, $\Lambda/\sqrt{|c_{HD}|}$, and $\Lambda/\sqrt{|c_{H\Box}|}$ as a function of the central values of $\Lambda/\sqrt{|c_H|}$, respectively.
We assumed the central values in the left panels of Figs.~\ref{fig:sens1}-\ref{fig:sens3} as $c_{uH}=c_{HD}=c_{H\Box}=0$.
The black and red curves correspond to the DECIGO and BBO experiments with $T_{\rm obs} =$ 1-year statistics and $v_b=0.3$, respectively.
Each point denotes the magnitude of 95~\% confidence intervals for $\Lambda/\sqrt{c_{uH}}$, $\Lambda/\sqrt{c_{HD}}$, and $\Lambda/\sqrt{c_{H\Box}}$ directions. 
The dotted blue lines are the sensitivity reach of the ILC-250 at 95\%~C.L.~\cite{deBlas:2022ofj,DeBlas:2019qco}, and the dotted red lines are the current bounds at 95\%~C.L.~\cite{Ethier:2021bye}.
Comparing Figs.~\ref{fig:SNRCuH}-\ref{fig:SNRCHBox} with Figs.~\ref{fig:sens1}-\ref{fig:sens3}, the sensitivity reach of the DECIGO and BBO experiments denotes $\sim 1$ TeV for SNR $\sim 10$, which denotes a typical value for the precisely measurable $c_{uH}$, $c_{H\Box}$, and $c_{HD}$.
It is found that the sensitivity reach of the DECIGO and BBO experiments may exceed that of future collider experiments when the SFO-EWPT arises by SMEFT $\mathcal{O}_H$ operators.

Lastly, we comment on the uncertainty effects of the coefficient $\Lambda/\sqrt{c_H}$ on the potential sensitivities to the Wilson coefficients $c_{uH}$, $c_{HD}$, and $c_{H\Box}$.
The potential sensitivities to the Wilson coefficients $c_{uH}$, $c_{HD}$, and $c_{H\Box}$ in the left panels of Figs.~\ref{fig:sens1}-\ref{fig:CHBoxvbd} are estimated by changing the central values of the coefficient $\Lambda/\sqrt{c_H}$, and we can regard the shift of points along the horizontal axes as the uncertainties coming from the coefficient $\Lambda/\sqrt{c_H}$. 
On the other hand, as shown in the right panels of Figs.~\ref{fig:sens1}-\ref{fig:CHBoxvbd}, the uncertainties of the coefficient $\Lambda/\sqrt{c_H}$ in the GW observations are $\mathcal{O}(10^{-3})$ TeV.
Even if each point of the left panels of Figs.~\ref{fig:sens1}-\ref{fig:CHBoxvbd} change along the horizontal axes within this uncertainties, i.e., $\mathcal{O}(10^{-3})$ TeV, the potential sensitivity to the Wilson coefficients $c_{uH}$, $c_{HD}$, and $c_{H\Box}$ does not change much.
Therefore, the sensitivity to the Wilson coefficients would not change much even when the uncertainty of the coefficient $c_H$ in the GW observations is taken into account.
However, we did not consider the theoretical uncertainty of the coefficient $c_H$. 
As shown in Ref.~\cite{Croon:2020cgk}, large theoretical uncertainties in the peak gravitational wave amplitude due to renormalization scale dependence have been pointed out.  
To take into account the theoretical uncertainties in the sensitivity to the Wilson coefficients $c_{uH}$, $c_{HD}$, and $c_{H\Box}$, we have to perform the same analysis as this work by changing the renormalization scale.
In the future, such analysis would be necessary.

\section{Discussion}
\label{sec:Dis}
%
As studied in Refs.~\cite{Damgaard:2015con,Postma:2020toi}, the SMEFT dimension-six operator descriptions of the SFO-EWPT are questioned\footnote{In Ref.~\cite{deVries:2017ncy}, the validity of the SMEFT dimension-six operators descriptions is also studied by focusing on the electroweak baryogenesis.}.
In Ref.~\cite{Damgaard:2015con}, the validity of the SMEFT is investigated by performing comparisons between the SFO-EWPT in the SM with a singlet scalar boson and $\mathcal{O}_H$ operator, and it is shown that its validity is limited to a small parameter space of the full singlet model.
Also, in Ref.~\cite{Postma:2020toi}, its validity is studied in a nearly model-independent way based on the covariant derivative expansion, and it is shown that the SFO-EWPT in a wide range of theories cannot be described by the SMEFT truncated up to dimension-six operators.
An exception to this argument of Ref.~\cite{Postma:2020toi} is the singlet-extended SM\footnote{The SM with multiple singlet fields is also included in this exception.}, and agreement of the SFO-EWPT in the SMEFT with the UV theory is possible if dimension-eight operators are included.
So far, we have investigated the sensitivity reach of the future GW observations to the SMEFT dimension-six operators focusing on the scenario that the SFO-EWPT arises from the dimension-six $\mathcal{O}_H$ operator. 
However, similar analysis as the above sections works well once the SFO-EWPT arises from the new physics effects.
To clarify this point, we provide the sensitivity reaches of the future gravitational observations by adding the SMEFT dimension-eight $\varphi^8$ operator to Eq.~\eqref{eq:HiggsP}. 
As discussed above, the SMEFT with an additional dimension-eight $\varphi^8$ operator can be UV completed in a singlet extended SM and describe the SFO-EWPT.
Instead of Eq.~\eqref{eq:HiggsP}, we consider the following Higgs potential:
\begin{align}
V\to V+ \frac{C_H^2}{16}\cdot\varphi^8.
\end{align}
The confidence intervals in future GW experiments for the SMEFT Wilson coefficients are shown in Figs.~\ref{fig:sens1Dim8}-\ref{fig:sens3Dim8}.
It is found that the future gravitational wave observations are potentially sensitive to the subleading corrections to the Higgs potential, regardless of whether the SFO-EWPT arises only from the dimension-six $\mathcal{O}_H$ operator.
Although we considered the SMEFT operators for the SFO-EWPT, the analyses of this work are applicable even if the SFO-EWPT is achieved by a concrete UV model, and potential sensitivity reaches of the future GW measurement to the SMEFT dimension-six operator effects can be derived.

\section{Summary}
\label{sec:summ}
We studied the SMEFT dimension-six operator effects on the spectrum of GWs produced from the SFO-EWPT.
The three types of the SMEFT operator effects, i.e., (i) $\mathcal{O}_H$ operator tree level effects, (ii) $\mathcal{O}_{H\Box}$ and $\mathcal{O}_{HD}$ operators tree level effects on the wave function renormalization of the Higgs field, and (iii) $\mathcal{O}_{uH}$ operator one-loop level effects were considered.
Firstly, we provided formulae of the SMEFT effects on the Higgs potential and calculated $v_c/T_c$ as functions of the three types of SMEFT operators, as shown in Figs.~\ref{fig:VCTC}-\ref{fig:VCTCHBox}.
We focused on the scenario where the GWs mainly arises by (i), and the GW spectrum is slightly shifted by (ii) and (iii) because (i) is the dominant effect on the SFO-EWPT that can produce the GWs; see Figs.~\ref{fig:VCTC}-\ref{fig:GWsp}. 
We numerically evaluated the sensitivity of the future GW observations to (ii) and (iii) by performing the Fisher matrix analysis.
Finally, the sensitivity of the future GW observations to (ii) and (iii) was compared with that of the future collider experiments. 

\begin{figure*}[t]
\centering
\includegraphics[width=0.45\textwidth]{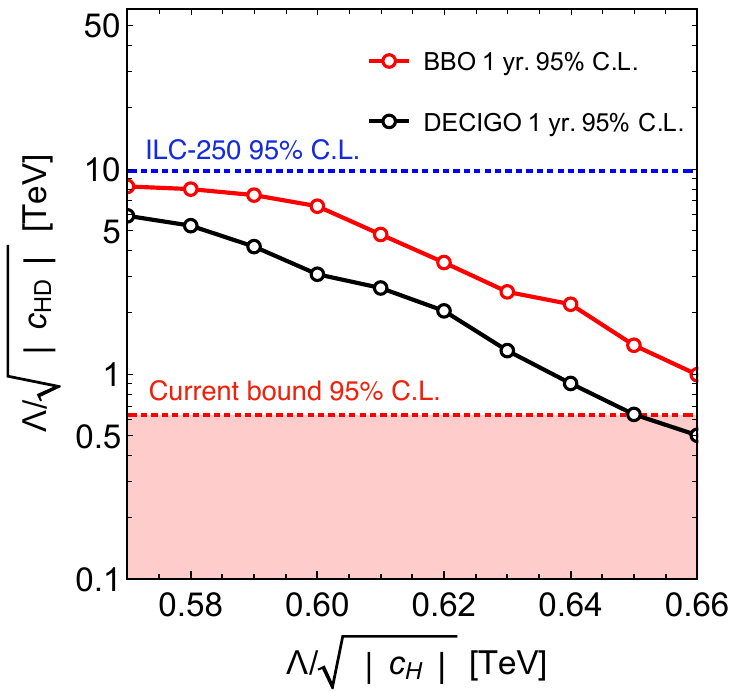}
\includegraphics[width=0.46\textwidth]{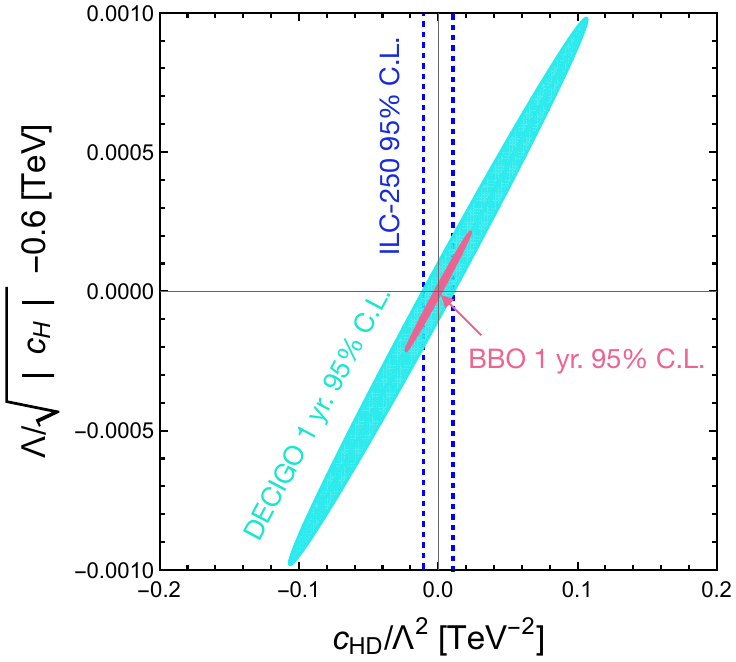}
\caption{
The same plots as Fig.~\ref{fig:sens1} but for $c_{HD}/\Lambda^2$.
\label{fig:sens2}
}
\end{figure*}

The results are collected in Figs.~\ref{fig:SNRCuH}-\ref{fig:sens3}.
We found that the DECIGO and the BBO experiments can be sensitive to (ii) and (iii) once the SFO-EWPT arises from (i).
In particular, its sensitivities to the operators $\mathcal{O}_{uH}$ and $\mathcal{O}_{H\Box}$ are potentially higher than future collider experiments.
Also, these results hold even if the SFO-EWPT arises from the SMEFT dimension-eight operator in addition to (i).
When the central value of $C_H$ is determined from the collider experiments, future GW observations potentially measure the SMEFT Wilson coefficients with high precision and yield constraints on the SMEFT complementary to future collider experiments.

\appendix

\section{Sensitivity reach of DECIGO at different bench mark point of bubble wall velocity}
\label{app:vel}
\begin{figure*}[t]
\centering
\includegraphics[width=0.45\textwidth]{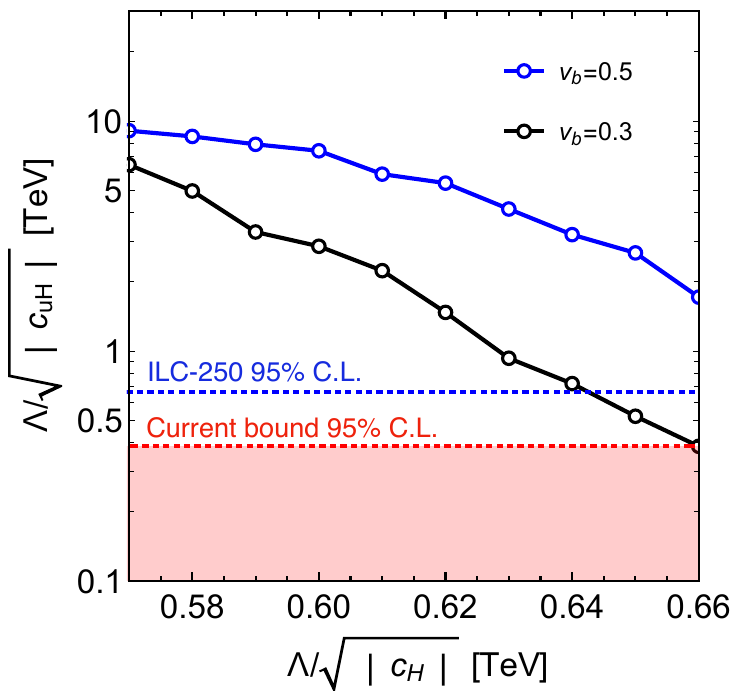}
\includegraphics[width=0.46\textwidth]{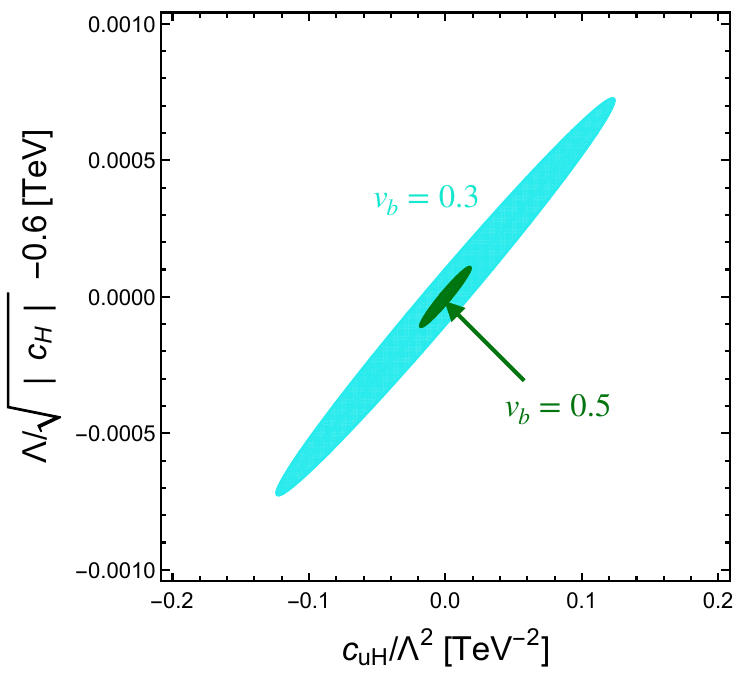}
\caption{
Left panel: Sensitivity reach of DECIGO with $T_{\rm obs}=$ 1-year statistics to $\Lambda/\sqrt{|c_{uH}|}$ as a function of the central value of $\Lambda/\sqrt{|c_H|}$, assuming the central value of $c_{uH}=0$.
The black and blue curves represent $v_b=0.3$ and $0.5$, respectively.
Each point denotes the magnitude of confidence intervals of $\Lambda/\sqrt{|c_{uH}|}$ at 95 \% C.L. ,i.e., $\delta \chi^2=6.0$, depending on the central values of $\Lambda/\sqrt{|c_H|}$.
The dotted red line is the current bound at 95\% C.L.~\cite{Ethier:2021bye}.
The dashed blue line shows a sensitivity reach of the ILC-250 at 95\% C.L.~\cite{DeBlas:2019qco,deBlas:2022ofj}.
Right panel: 95\% C.L. confidence regions for DECIGO with $T_{\rm obs}=$ 1-year statistics for $v_b=0.3$ (shaded blue) and $0.5$ (shaded green), assuming the central values of $c_{uH}/\Lambda^2=0$ and $\Lambda/\sqrt{|c_H|}=0.60~{\rm TeV}$.
The confidence intervals for $c_{uH}/\Lambda^2$ direction in the right panel correspond to one of points in the left panel.
\label{fig:CuHvbd}
}
\end{figure*}

\begin{figure*}[t]
\centering
\includegraphics[width=0.45\textwidth]{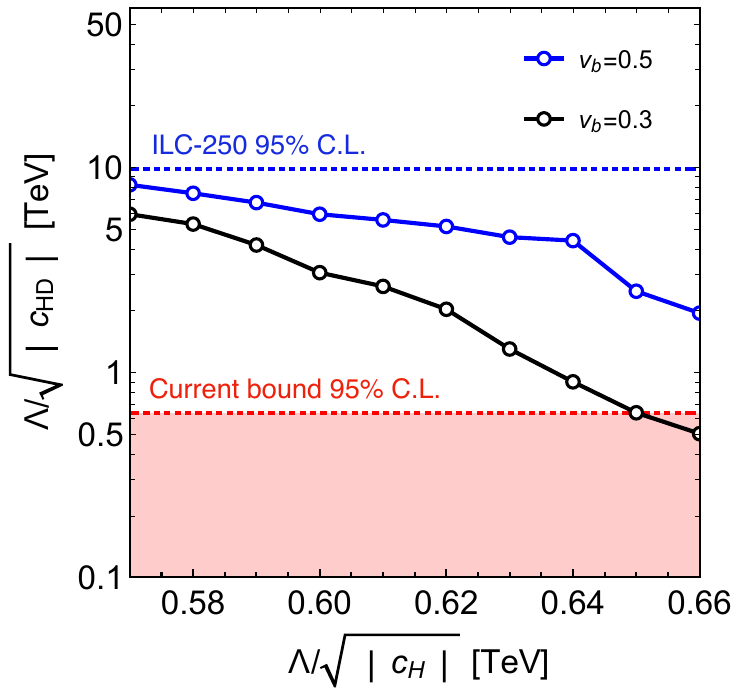}
\includegraphics[width=0.46\textwidth]{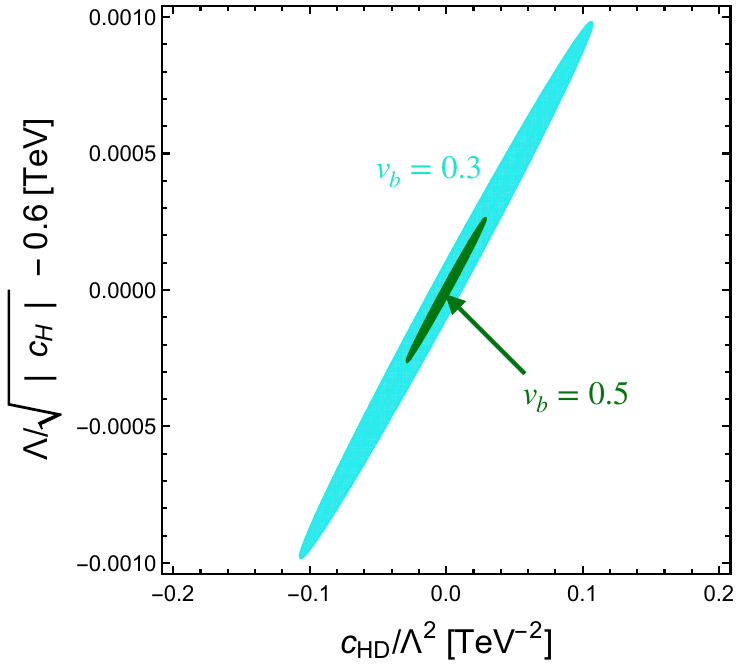}
\caption{
The same plots as Fig.~\ref{fig:CuHvbd} but for $c_{HD}/\Lambda^2$.
\label{fig:CHDvbd}
}
\end{figure*}

\begin{figure*}[t]
\centering
\includegraphics[width=0.45\textwidth]{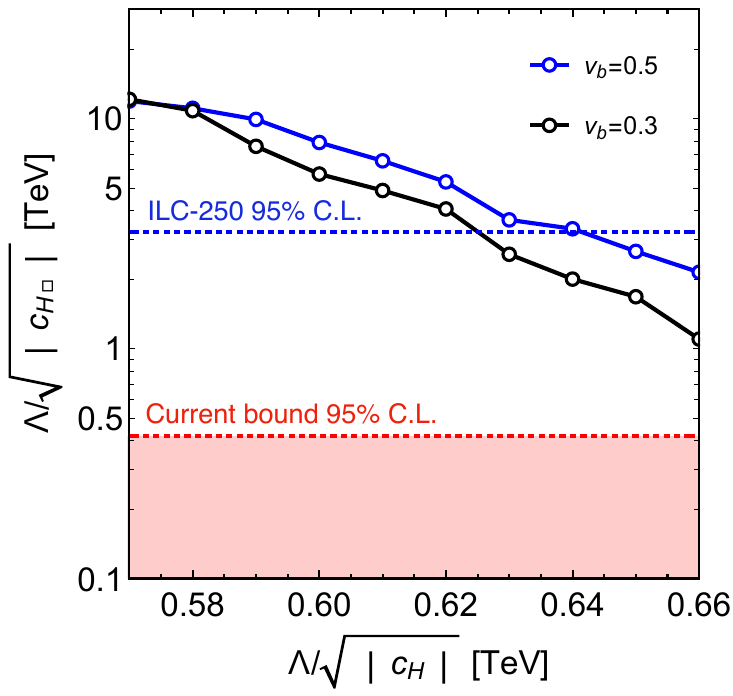}
\includegraphics[width=0.46\textwidth]{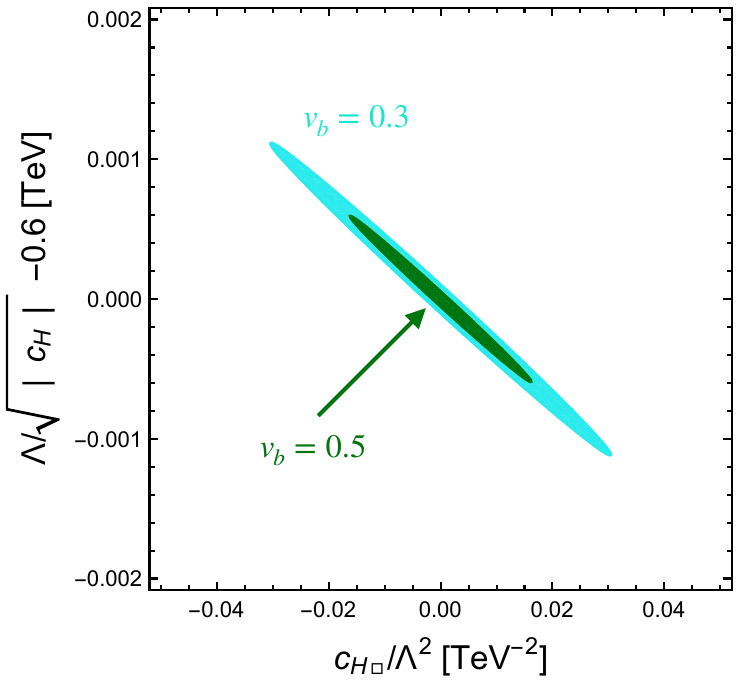}
\caption{
The same plots as Fig.~\ref{fig:CuHvbd} but for $c_{H\Box}/\Lambda^2$.
\label{fig:CHBoxvbd}
}
\end{figure*}

In Figs.~\ref{fig:CuHvbd}-\ref{fig:CHBoxvbd}, the results of DECIGO with $T_{\rm obs}=$ 1-year statistics are shown in the case that the benchmark point of the bubble wall velocity is $0.3$, and $0.5$.
Figs.~\ref{fig:CuHvbd}-\ref{fig:CHBoxvbd} show that the GW spectrum increases as the bubble wall velocity increases.

\begin{figure*}[t]
\centering
\includegraphics[width=0.45\textwidth]{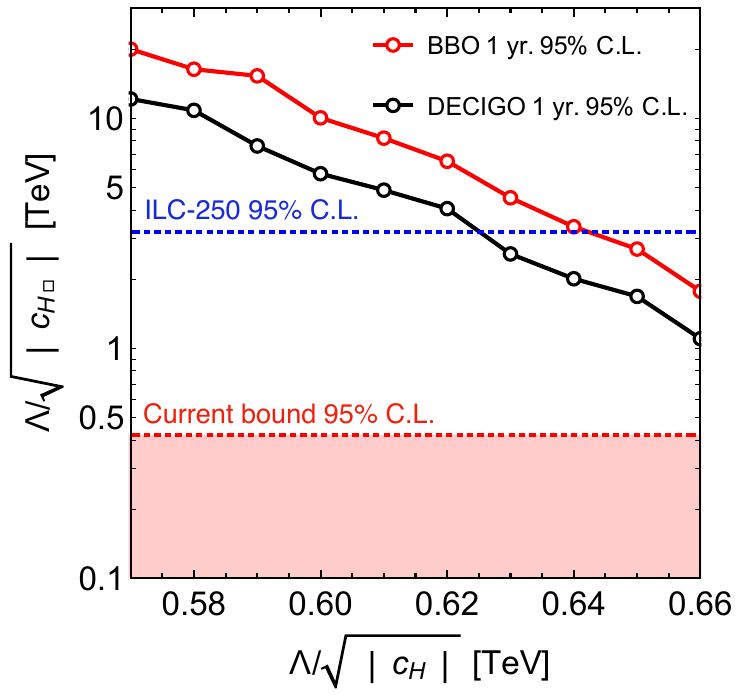}
\includegraphics[width=0.46\textwidth]{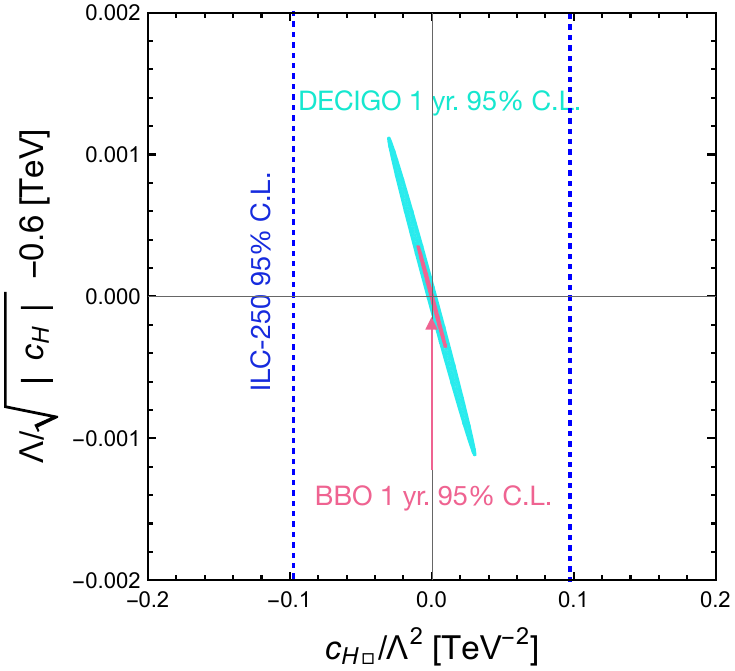}
\caption{
The same plots as Fig.~\ref{fig:sens1} but for $c_{H\Box}/\Lambda^2$.
\label{fig:sens3}
}
\end{figure*}

\begin{figure*}[t]
\centering
\includegraphics[width=0.45\textwidth]{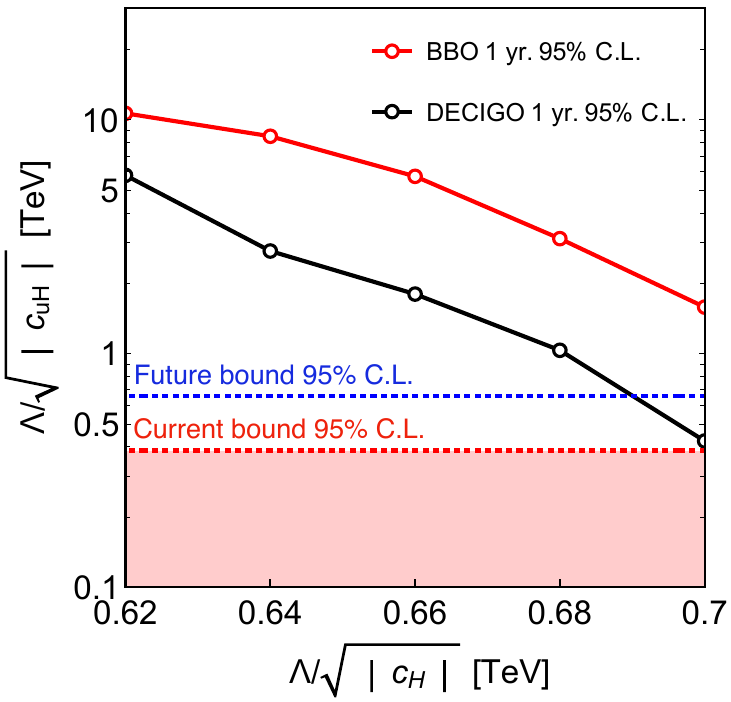}
\includegraphics[width=0.46\textwidth]{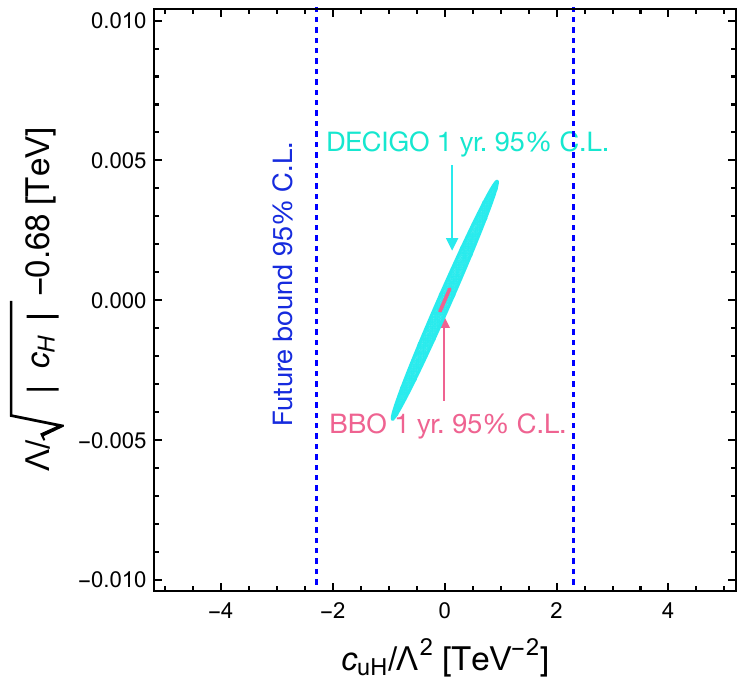}
\caption{
Sensitivity reach of the DECIGO and BBO experiments to the SMEFT effects when the SFO-EWPT is achieved by the SMEFT dimension-eight $\varphi^8$ operator in addition to $\mathcal{O}_H$.
Left panel: Sensitivity reach of the DECIGO (black) and BBO (red) experiments to $\Lambda/\sqrt{|c_{uH}|}$ as a function of the central value of $\Lambda/\sqrt{|c_H|}$, assuming $v_b=0.3$, the central value of $c_{uH}/\Lambda^2=0$, and $T_{\rm obs} = $ 1-year statistics.
The vertical axis of the left panel is not the central value of $\Lambda/\sqrt{|c_{uH}|}$, but the magnitude of 95\% confidence intervals of $\Lambda/\sqrt{|c_{uH}|}$.
The red-dotted line is the current bound at 95\% C.L.~\cite{Ethier:2021bye}.
The dashed blue line shows the future bound at 95\% C.L.~\cite{DeBlas:2019qco,deBlas:2022ofj}.
Right panel: 95\% C.L. confidence regions for DECIGO (blue-shaded) and BBO (red-shaded) with 1-year statistics, assuming the central values of $c_{uH}/\Lambda^2=0$ and $\Lambda/\sqrt{|c_H|}=0.68~{\rm TeV}$.
The confidence interval for $c_{uH}/\Lambda^2$ direction in the right panel corresponds to one of the points in the left panel.}
\label{fig:sens1Dim8}
\end{figure*}

\begin{figure*}[t]
\centering
\includegraphics[width=0.45\textwidth]{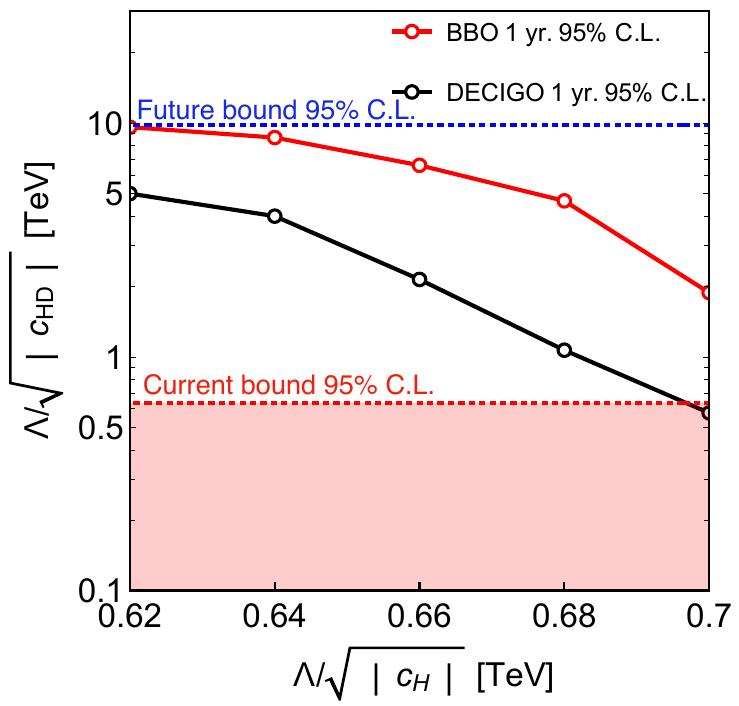}
\includegraphics[width=0.46\textwidth]{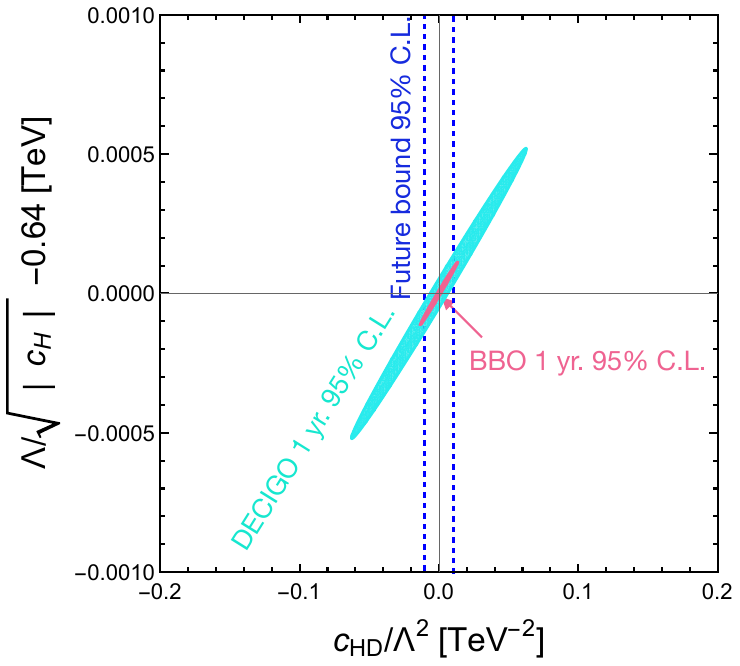}
\caption{
The same plots as Fig.~\ref{fig:sens1Dim8} but for $c_{HD}/\Lambda^2$.
\label{fig:sens2Dim8}
}
\end{figure*}

\begin{figure*}[t]
\centering
\includegraphics[width=0.45\textwidth]{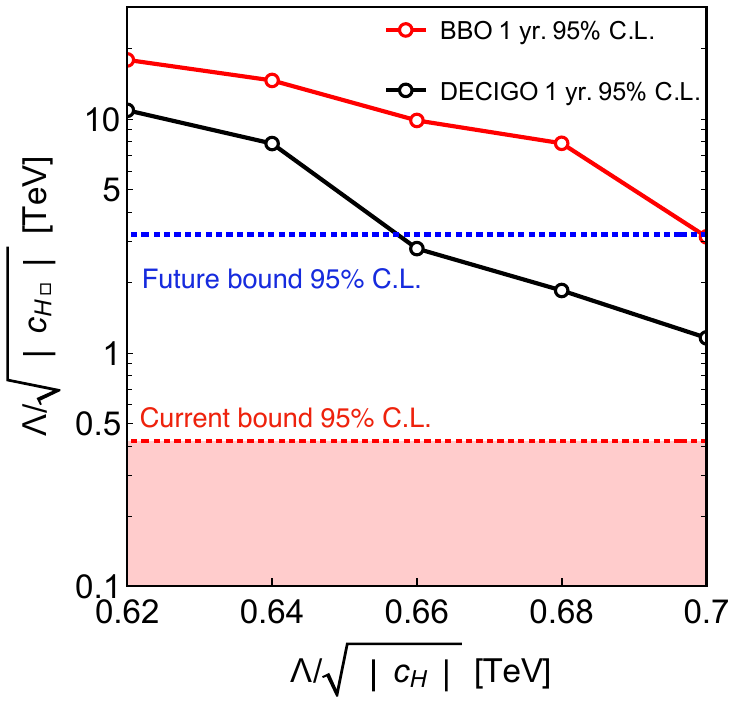}
\includegraphics[width=0.46\textwidth]{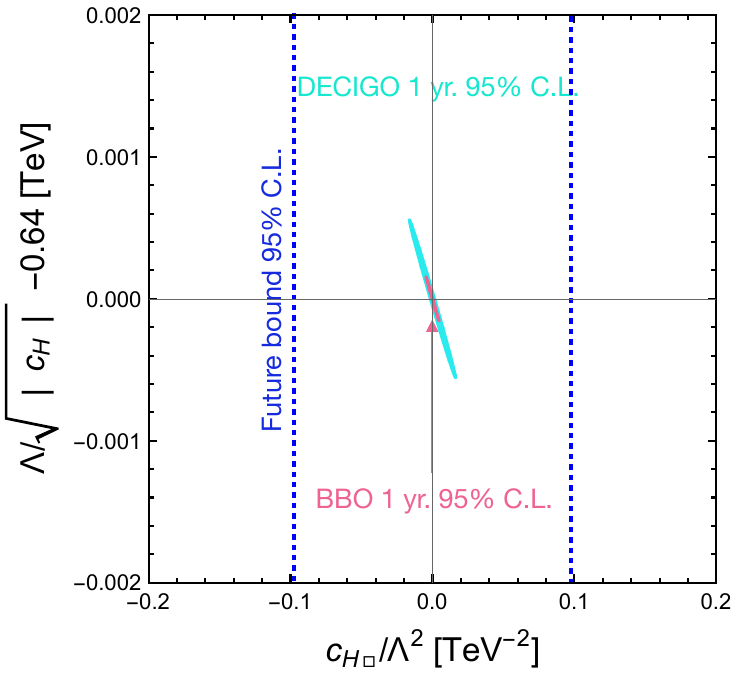}
\caption{
The same plots as Fig.~\ref{fig:sens1Dim8} but for $c_{H\Box}/\Lambda^2$.
\label{fig:sens3Dim8}
}
\end{figure*}

\clearpage
\bibliography{SMEFTPT.bib}

\providecommand{\href}[2]{#2}\begingroup\begin{thebibliography}{10}

\bibitem{ATLAS:2012yve}
{\bfseries ATLAS} Collaboration, ``{Observation of a new particle in the search
  for the Standard Model Higgs boson with the ATLAS detector at the LHC},''
  \href{https://dx.doi.org/10.1016/j.physletb.2012.08.020}{Phys.\  Lett.\  B
  {\bfseries 716} (2012) 1--29} {\ttfamily
  [\href{https://arxiv.org/abs/1207.7214}{arXiv:1207.7214}]}.

\bibitem{CMS:2012qbp}
{\bfseries CMS} Collaboration, ``{Observation of a New Boson at a Mass of 125
  GeV with the CMS Experiment at the LHC},''
  \href{https://dx.doi.org/10.1016/j.physletb.2012.08.021}{Phys.\  Lett.\  B
  {\bfseries 716} (2012) 30--61} {\ttfamily
  [\href{https://arxiv.org/abs/1207.7235}{arXiv:1207.7235}]}.

\bibitem{Gavela:1993ts}
M.~B.~Gavela, P.~Hernandez, J.~Orloff, and O.~Pene, ``{Standard model CP
  violation and baryon asymmetry},''
  \href{https://dx.doi.org/10.1142/S0217732394000629}{Mod.\  Phys.\  Lett.\  A
  {\bfseries 9} (1994) 795--810} {\ttfamily
  [\href{https://arxiv.org/abs/hep-ph/9312215}{hep-ph/9312215}]}.

\bibitem{Konstandin:2003dx}
T.~Konstandin, T.~Prokopec, and M.~G.~Schmidt, ``{Axial currents from CKM
  matrix CP violation and electroweak baryogenesis},''
  \href{https://dx.doi.org/10.1016/j.nuclphysb.2003.11.037}{Nucl.\  Phys.\  B
  {\bfseries 679} (2004) 246--260} {\ttfamily
  [\href{https://arxiv.org/abs/hep-ph/0309291}{hep-ph/0309291}]}.

\bibitem{Kuzmin:1985mm}
V.~A.~Kuzmin, V.~A.~Rubakov, and M.~E.~Shaposhnikov, ``{On the Anomalous
  Electroweak Baryon Number Nonconservation in the Early Universe},''
  \href{https://dx.doi.org/10.1016/0370-2693(85)91028-7}{Phys.\  Lett.\  B
  {\bfseries 155} (1985) 36}.

\bibitem{Kajantie:1995kf}
K.~Kajantie, M.~Laine, K.~Rummukainen, and M.~E.~Shaposhnikov, ``{The
  Electroweak phase transition: A Nonperturbative analysis},''
  \href{https://dx.doi.org/10.1016/0550-3213(96)00052-1}{Nucl.\  Phys.\  B
  {\bfseries 466} (1996) 189--258} {\ttfamily
  [\href{https://arxiv.org/abs/hep-lat/9510020}{hep-lat/9510020}]}.

\bibitem{Kajantie:1996mn}
K.~Kajantie, M.~Laine, K.~Rummukainen, and M.~E.~Shaposhnikov, ``{Is there a~
  hot electroweak phase transition at $m_H \gtrsim m_W$?}''
  \href{https://dx.doi.org/10.1103/PhysRevLett.77.2887}{Phys.\  Rev.\  Lett.\
  {\bfseries 77} (1996) 2887--2890} {\ttfamily
  [\href{https://arxiv.org/abs/hep-ph/9605288}{hep-ph/9605288}]}.

\bibitem{Kajantie:1996qd}
K.~Kajantie, M.~Laine, K.~Rummukainen, and M.~E.~Shaposhnikov, ``{A
  Nonperturbative analysis of the finite T phase transition in SU(2) x U(1)
  electroweak theory},''
  \href{https://dx.doi.org/10.1016/S0550-3213(97)00164-8}{Nucl.\  Phys.\  B
  {\bfseries 493} (1997) 413--438} {\ttfamily
  [\href{https://arxiv.org/abs/hep-lat/9612006}{hep-lat/9612006}]}.

\bibitem{Csikor:1998eu}
F.~Csikor, Z.~Fodor, and J.~Heitger, ``{Endpoint of the hot electroweak phase
  transition},'' \href{https://dx.doi.org/10.1103/PhysRevLett.82.21}{Phys.\
  Rev.\  Lett.\  {\bfseries 82} (1999) 21--24} {\ttfamily
  [\href{https://arxiv.org/abs/hep-ph/9809291}{hep-ph/9809291}]}.

\bibitem{DOnofrio:2015gop}
M.~D'Onofrio and K.~Rummukainen, ``{Standard model cross-over on the
  lattice},'' \href{https://dx.doi.org/10.1103/PhysRevD.93.025003}{Phys.\
  Rev.\  D {\bfseries 93} (2016) 025003} {\ttfamily
  [\href{https://arxiv.org/abs/1508.07161}{arXiv:1508.07161}]}.

\bibitem{Workman:2022ynf}
{\bfseries Particle Data Group} Collaboration, ``{Review of Particle
  Physics},'' \href{https://dx.doi.org/10.1093/ptep/ptac097}{PTEP {\bfseries
  2022} (2022) 083C01}.

\bibitem{Grzadkowski:2010es}
B.~Grzadkowski, M.~Iskrzynski, M.~Misiak, and J.~Rosiek, ``{Dimension-Six Terms
  in the Standard Model Lagrangian},''
  \href{https://dx.doi.org/10.1007/JHEP10(2010)085}{JHEP {\bfseries 10} (2010)
  085} {\ttfamily [\href{https://arxiv.org/abs/1008.4884}{arXiv:1008.4884}]}.

\bibitem{Jenkins:2013zja}
E.~E.~Jenkins, A.~V.~Manohar, and M.~Trott, ``{Renormalization Group Evolution
  of the Standard Model Dimension Six Operators I: Formalism and lambda
  Dependence},'' \href{https://dx.doi.org/10.1007/JHEP10(2013)087}{JHEP
  {\bfseries 10} (2013) 087} {\ttfamily
  [\href{https://arxiv.org/abs/1308.2627}{arXiv:1308.2627}]}.

\bibitem{Jenkins:2013wua}
E.~E.~Jenkins, A.~V.~Manohar, and M.~Trott, ``{Renormalization Group Evolution
  of the Standard Model Dimension Six Operators II: Yukawa Dependence},''
  \href{https://dx.doi.org/10.1007/JHEP01(2014)035}{JHEP {\bfseries 01} (2014)
  035} {\ttfamily [\href{https://arxiv.org/abs/1310.4838}{arXiv:1310.4838}]}.

\bibitem{Alonso:2013hga}
R.~Alonso, E.~E.~Jenkins, A.~V.~Manohar, and M.~Trott, ``{Renormalization Group
  Evolution of the Standard Model Dimension Six Operators III: Gauge Coupling
  Dependence and Phenomenology},''
  \href{https://dx.doi.org/10.1007/JHEP04(2014)159}{JHEP {\bfseries 04} (2014)
  159} {\ttfamily [\href{https://arxiv.org/abs/1312.2014}{arXiv:1312.2014}]}.

\bibitem{Grojean:2004xa}
C.~Grojean, G.~Servant, and J.~D.~Wells, ``{First-order electroweak phase
  transition in the standard model with a low cutoff},''
  \href{https://dx.doi.org/10.1103/PhysRevD.71.036001}{Phys.\  Rev.\  D
  {\bfseries 71} (2005) 036001} {\ttfamily
  [\href{https://arxiv.org/abs/hep-ph/0407019}{hep-ph/0407019}]}.

\bibitem{Zhang:1992fs}
X.-m.~Zhang, ``{Operators analysis for Higgs potential and cosmological bound
  on Higgs mass},'' \href{https://dx.doi.org/10.1103/PhysRevD.47.3065}{Phys.\
  Rev.\  D {\bfseries 47} (1993) 3065--3067} {\ttfamily
  [\href{https://arxiv.org/abs/hep-ph/9301277}{hep-ph/9301277}]}.

\bibitem{Bodeker:2004ws}
D.~Bodeker, L.~Fromme, S.~J.~Huber, and M.~Seniuch, ``{The Baryon asymmetry in
  the standard model with a low cut-off},''
  \href{https://dx.doi.org/10.1088/1126-6708/2005/02/026}{JHEP {\bfseries 02}
  (2005) 026} {\ttfamily
  [\href{https://arxiv.org/abs/hep-ph/0412366}{hep-ph/0412366}]}.

\bibitem{Huber:2007vva}
S.~J.~Huber and T.~Konstandin, ``{Production of gravitational waves in the
  nMSSM},'' \href{https://dx.doi.org/10.1088/1475-7516/2008/05/017}{JCAP
  {\bfseries 05} (2008) 017} {\ttfamily
  [\href{https://arxiv.org/abs/0709.2091}{arXiv:0709.2091}]}.

\bibitem{Delaunay:2007wb}
C.~Delaunay, C.~Grojean, and J.~D.~Wells, ``{Dynamics of Non-renormalizable
  Electroweak Symmetry Breaking},''
  \href{https://dx.doi.org/10.1088/1126-6708/2008/04/029}{JHEP {\bfseries 04}
  (2008) 029} {\ttfamily
  [\href{https://arxiv.org/abs/0711.2511}{arXiv:0711.2511}]}.

\bibitem{Huber:2013kj}
S.~J.~Huber and M.~Sopena, ``{An efficient approach to electroweak bubble
  velocities}.'' {\ttfamily
  \href{https://arxiv.org/abs/1302.1044}{arXiv:1302.1044}}.

\bibitem{Konstandin:2014zta}
T.~Konstandin, G.~Nardini, and I.~Rues, ``{From Boltzmann equations to steady
  wall velocities},''
  \href{https://dx.doi.org/10.1088/1475-7516/2014/09/028}{JCAP {\bfseries 09}
  (2014) 028} {\ttfamily
  [\href{https://arxiv.org/abs/1407.3132}{arXiv:1407.3132}]}.

\bibitem{Damgaard:2015con}
P.~H.~Damgaard, A.~Haarr, D.~O'Connell, and A.~Tranberg, ``{Effective Field
  Theory and Electroweak Baryogenesis in the Singlet-Extended Standard
  Model},'' \href{https://dx.doi.org/10.1007/JHEP02(2016)107}{JHEP {\bfseries
  02} (2016) 107} {\ttfamily
  [\href{https://arxiv.org/abs/1512.01963}{arXiv:1512.01963}]}.

\bibitem{Harman:2015gif}
C.~P.~D.~Harman and S.~J.~Huber, ``{Does zero temperature decide on the nature
  of the electroweak phase transition?}''
  \href{https://dx.doi.org/10.1007/JHEP06(2016)005}{JHEP {\bfseries 06} (2016)
  005} {\ttfamily [\href{https://arxiv.org/abs/1512.05611}{arXiv:1512.05611}]}.

\bibitem{Balazs:2016yvi}
C.~Balazs, G.~White, and J.~Yue, ``{Effective field theory, electric dipole
  moments and electroweak baryogenesis},''
  \href{https://dx.doi.org/10.1007/JHEP03(2017)030}{JHEP {\bfseries 03} (2017)
  030} {\ttfamily [\href{https://arxiv.org/abs/1612.01270}{arXiv:1612.01270}]}.

\bibitem{deVries:2017ncy}
J.~de~Vries, M.~Postma, J.~van~de Vis, and G.~White, ``{Electroweak
  Baryogenesis and the Standard Model Effective Field Theory},''
  \href{https://dx.doi.org/10.1007/JHEP01(2018)089}{JHEP {\bfseries 01} (2018)
  089} {\ttfamily [\href{https://arxiv.org/abs/1710.04061}{arXiv:1710.04061}]}.

\bibitem{Cai:2017tmh}
R.-G.~Cai, M.~Sasaki, and S.-J.~Wang, ``{The gravitational waves from the
  first-order phase transition with a dimension-six operator},''
  \href{https://dx.doi.org/10.1088/1475-7516/2017/08/004}{JCAP {\bfseries 08}
  (2017) 004} {\ttfamily
  [\href{https://arxiv.org/abs/1707.03001}{arXiv:1707.03001}]}.

\bibitem{Chala:2018ari}
M.~Chala, C.~Krause, and G.~Nardini, ``{Signals of the electroweak phase
  transition at colliders and gravitational wave observatories},''
  \href{https://dx.doi.org/10.1007/JHEP07(2018)062}{JHEP {\bfseries 07} (2018)
  062} {\ttfamily [\href{https://arxiv.org/abs/1802.02168}{arXiv:1802.02168}]}.

\bibitem{Dorsch:2018pat}
G.~C.~Dorsch, S.~J.~Huber, and T.~Konstandin, ``{Bubble wall velocities in the
  Standard Model and beyond},''
  \href{https://dx.doi.org/10.1088/1475-7516/2018/12/034}{JCAP {\bfseries 12}
  (2018) 034} {\ttfamily
  [\href{https://arxiv.org/abs/1809.04907}{arXiv:1809.04907}]}.

\bibitem{DeVries:2018aul}
J.~De~Vries, M.~Postma, and J.~van~de Vis, ``{The role of leptons in
  electroweak baryogenesis},''
  \href{https://dx.doi.org/10.1007/JHEP04(2019)024}{JHEP {\bfseries 04} (2019)
  024} {\ttfamily [\href{https://arxiv.org/abs/1811.11104}{arXiv:1811.11104}]}.

\bibitem{Chala:2019rfk}
M.~Chala, V.~V.~Khoze, M.~Spannowsky, and P.~Waite, ``{Mapping the shape of the
  scalar potential with gravitational waves},''
  \href{https://dx.doi.org/10.1142/S0217751X19502233}{Int.\  J.\  Mod.\  Phys.\
   A {\bfseries 34} (2019) 1950223} {\ttfamily
  [\href{https://arxiv.org/abs/1905.00911}{arXiv:1905.00911}]}.

\bibitem{Ellis:2019flb}
S.~A.~R.~Ellis, S.~Ipek, and G.~White, ``{Electroweak Baryogenesis from
  Temperature-Varying Couplings},''
  \href{https://dx.doi.org/10.1007/JHEP08(2019)002}{JHEP {\bfseries 08} (2019)
  002} {\ttfamily [\href{https://arxiv.org/abs/1905.11994}{arXiv:1905.11994}]}.

\bibitem{Zhou:2019uzq}
R.~Zhou, L.~Bian, and H.-K.~Guo, ``{Connecting the electroweak sphaleron with
  gravitational waves},''
  \href{https://dx.doi.org/10.1103/PhysRevD.101.091903}{Phys.\  Rev.\  D
  {\bfseries 101} (2020) 091903} {\ttfamily
  [\href{https://arxiv.org/abs/1910.00234}{arXiv:1910.00234}]}.

\bibitem{Kanemura:2020yyr}
S.~Kanemura and M.~Tanaka, ``{Higgs boson coupling as a probe of the sphaleron
  property},'' \href{https://dx.doi.org/10.1016/j.physletb.2020.135711}{Phys.\
  Lett.\  B {\bfseries 809} (2020) 135711} {\ttfamily
  [\href{https://arxiv.org/abs/2005.05250}{arXiv:2005.05250}]}.

\bibitem{Phong:2020ybr}
V.~Q.~Phong, P.~H.~Khiem, N.~P.~D.~Loc, and H.~N.~Long, ``{Sphaleron in the
  first-order electroweak phase transition with the dimension-six Higgs field
  operator},'' \href{https://dx.doi.org/10.1103/PhysRevD.101.116010}{Phys.\
  Rev.\  D {\bfseries 101} (2020) 116010} {\ttfamily
  [\href{https://arxiv.org/abs/2003.09625}{arXiv:2003.09625}]}.

\bibitem{Wang:2020zlf}
X.~Wang, F.~P.~Huang, and X.~Zhang, ``{Bubble wall velocity beyond leading-log
  approximation in electroweak phase transition}.'' {\ttfamily
  \href{https://arxiv.org/abs/2011.12903}{arXiv:2011.12903}}.

\bibitem{Wang:2020jrd}
X.~Wang, F.~P.~Huang, and X.~Zhang, ``{Phase transition dynamics and
  gravitational wave spectra of strong first-order phase transition in
  supercooled universe},''
  \href{https://dx.doi.org/10.1088/1475-7516/2020/05/045}{JCAP {\bfseries 05}
  (2020) 045} {\ttfamily
  [\href{https://arxiv.org/abs/2003.08892}{arXiv:2003.08892}]}.

\bibitem{Kanemura:2021fvp}
S.~Kanemura and R.~Nagai, ``{A new Higgs effective field theory and the new
  no-lose theorem},'' \href{https://dx.doi.org/10.1007/JHEP03(2022)194}{JHEP
  {\bfseries 03} (2022) 194} {\ttfamily
  [\href{https://arxiv.org/abs/2111.12585}{arXiv:2111.12585}]}.

\bibitem{Lewicki:2021pgr}
M.~Lewicki, M.~Merchand, and M.~Zych, ``{Electroweak bubble wall expansion:
  gravitational waves and baryogenesis in Standard Model-like thermal
  plasma},'' \href{https://dx.doi.org/10.1007/JHEP02(2022)017}{JHEP {\bfseries
  02} (2022) 017} {\ttfamily
  [\href{https://arxiv.org/abs/2111.02393}{arXiv:2111.02393}]}.

\bibitem{Hashino:2021qoq}
K.~Hashino, S.~Kanemura, and T.~Takahashi, ``{Primordial black holes as a probe
  of strongly first-order electroweak phase transition},''
  \href{https://dx.doi.org/10.1016/j.physletb.2022.137261}{Phys.\  Lett.\  B
  {\bfseries 833} (2022) 137261} {\ttfamily
  [\href{https://arxiv.org/abs/2111.13099}{arXiv:2111.13099}]}.

\bibitem{Kanemura:2022txx}
S.~Kanemura, R.~Nagai, and M.~Tanaka, ``{Electroweak phase transition in the
  nearly aligned Higgs effective field theory},''
  \href{https://dx.doi.org/10.1007/JHEP06(2022)027}{JHEP {\bfseries 06} (2022)
  027} {\ttfamily [\href{https://arxiv.org/abs/2202.12774}{arXiv:2202.12774}]}.

\bibitem{Anisha:2022hgv}
Anisha, L.~Biermann, C.~Englert, and M.~M\"uhlleitner, ``{Two Higgs doublets,
  effective interactions and a strong first-order electroweak phase
  transition},'' \href{https://dx.doi.org/10.1007/JHEP08(2022)091}{JHEP
  {\bfseries 08} (2022) 091} {\ttfamily
  [\href{https://arxiv.org/abs/2204.06966}{arXiv:2204.06966}]}.

\bibitem{Croon:2020cgk}
D.~Croon, O.~Gould, P.~Schicho, T.~V.~I.~Tenkanen, and G.~White, ``{Theoretical
  uncertainties for cosmological first-order phase transitions},''
  \href{https://dx.doi.org/10.1007/JHEP04(2021)055}{JHEP {\bfseries 04} (2021)
  055} {\ttfamily [\href{https://arxiv.org/abs/2009.10080}{arXiv:2009.10080}]}.

\bibitem{Huang:2015izx}
F.~P.~Huang, P.-H.~Gu, P.-F.~Yin, Z.-H.~Yu, and X.~Zhang, ``{Testing the
  electroweak phase transition and electroweak baryogenesis at the LHC and a
  circular electron-positron collider},''
  \href{https://dx.doi.org/10.1103/PhysRevD.93.103515}{Phys.\  Rev.\  D
  {\bfseries 93} (2016) 103515} {\ttfamily
  [\href{https://arxiv.org/abs/1511.03969}{arXiv:1511.03969}]}.

\bibitem{Cao:2017oez}
Q.-H.~Cao, F.~P.~Huang, K.-P.~Xie, and X.~Zhang, ``{Testing the electroweak
  phase transition in scalar extension models at lepton colliders},''
  \href{https://dx.doi.org/10.1088/1674-1137/42/2/023103}{Chin.\  Phys.\  C
  {\bfseries 42} (2018) 023103} {\ttfamily
  [\href{https://arxiv.org/abs/1708.04737}{arXiv:1708.04737}]}.

\bibitem{Huang:2016odd}
F.~P.~Huang, Y.~Wan, D.-G.~Wang, Y.-F.~Cai, and X.~Zhang, ``{Hearing the echoes
  of electroweak baryogenesis with gravitational wave detectors},''
  \href{https://dx.doi.org/10.1103/PhysRevD.94.041702}{Phys.\  Rev.\  D
  {\bfseries 94} (2016) 041702} {\ttfamily
  [\href{https://arxiv.org/abs/1601.01640}{arXiv:1601.01640}]}.

\bibitem{Postma:2020toi}
M.~Postma and G.~White, ``{Cosmological phase transitions: is effective field
  theory just a toy?}'' \href{https://dx.doi.org/10.1007/JHEP03(2021)280}{JHEP
  {\bfseries 03} (2021) 280} {\ttfamily
  [\href{https://arxiv.org/abs/2012.03953}{arXiv:2012.03953}]}.

\bibitem{Huber:2006ri}
S.~J.~Huber, M.~Pospelov, and A.~Ritz, ``{Electric dipole moment constraints on
  minimal electroweak baryogenesis},''
  \href{https://dx.doi.org/10.1103/PhysRevD.75.036006}{Phys.\  Rev.\  D
  {\bfseries 75} (2007) 036006} {\ttfamily
  [\href{https://arxiv.org/abs/hep-ph/0610003}{hep-ph/0610003}]}.

\bibitem{Cepeda:2019klc}
A.~Dainese \emph{et al}., eds., ``{Report from Working Group 2}: {Higgs Physics
  at the HL-LHC and HE-LHC},''
  \href{https://dx.doi.org/10.23731/CYRM-2019-007.221}{CERN Yellow Rep.\
  Monogr.\  {\bfseries 7} (2019) 221--584} {\ttfamily
  [\href{https://arxiv.org/abs/1902.00134}{arXiv:1902.00134}]}.

\bibitem{LCCPhysicsWorkingGroup:2019fvj}
{\bfseries LCC Physics Working Group} Collaboration, ``{Tests of the Standard
  Model at the International Linear Collider}.'' {\ttfamily
  \href{https://arxiv.org/abs/1908.11299}{arXiv:1908.11299}}.

\bibitem{CLIC:2018fvx}
{\bfseries CLIC} Collaboration, ``{The CLIC Potential for New Physics}.''
  {\ttfamily \href{https://arxiv.org/abs/1812.02093}{arXiv:1812.02093}}.

\bibitem{FCC:2018byv}
{\bfseries FCC} Collaboration, ``{FCC Physics Opportunities}: {Future Circular
  Collider Conceptual Design Report Volume 1},''
  \href{https://dx.doi.org/10.1140/epjc/s10052-019-6904-3}{Eur.\  Phys.\  J.\
  C {\bfseries 79} (2019) 474}.

\bibitem{An:2018dwb}
F.~An \emph{et al}., ``{Precision Higgs physics at the CEPC},''
  \href{https://dx.doi.org/10.1088/1674-1137/43/4/043002}{Chin.\  Phys.\  C
  {\bfseries 43} (2019) 043002} {\ttfamily
  [\href{https://arxiv.org/abs/1810.09037}{arXiv:1810.09037}]}.

\bibitem{LISA:2017pwj}
{\bfseries LISA} Collaboration, ``{Laser Interferometer Space Antenna}.''
  {\ttfamily \href{https://arxiv.org/abs/1702.00786}{arXiv:1702.00786}}.

\bibitem{Seto:2005qy}
N.~Seto, ``{Correlation analysis of stochastic gravitational wave background
  around 0.1-1 Hz},''
  \href{https://dx.doi.org/10.1103/PhysRevD.73.063001}{Phys.\  Rev.\  D
  {\bfseries 73} (2006) 063001} {\ttfamily
  [\href{https://arxiv.org/abs/gr-qc/0510067}{gr-qc/0510067}]}.

\bibitem{BBO}
E.~S.~P.~{\em et~al. }, {The big bang observer: Direct detection of
  gravitational waves from the birth of the universe to the present,'' {\em
  NASA Mission Concept Study} (2004)} (unpublished).

\bibitem{Hashino:2018wee}
K.~Hashino, R.~Jinno, M.~Kakizaki, S.~Kanemura, \emph{et al}., ``{Selecting
  models of first-order phase transitions using the synergy between collider
  and gravitational-wave experiments},''
  \href{https://dx.doi.org/10.1103/PhysRevD.99.075011}{Phys.\  Rev.\  D
  {\bfseries 99} (2019) 075011} {\ttfamily
  [\href{https://arxiv.org/abs/1809.04994}{arXiv:1809.04994}]}.

\bibitem{Caprini:2019egz}
C.~Caprini \emph{et al}., ``{Detecting gravitational waves from cosmological
  phase transitions with LISA: an update},''
  \href{https://dx.doi.org/10.1088/1475-7516/2020/03/024}{JCAP {\bfseries 03}
  (2020) 024} {\ttfamily
  [\href{https://arxiv.org/abs/1910.13125}{arXiv:1910.13125}]}.

\bibitem{Hindmarsh:2017gnf}
M.~Hindmarsh, S.~J.~Huber, K.~Rummukainen, and D.~J.~Weir, ``{Shape of the
  acoustic gravitational wave power spectrum from a first order phase
  transition},'' \href{https://dx.doi.org/10.1103/PhysRevD.96.103520}{Phys.\
  Rev.\  D {\bfseries 96} (2017) 103520} {\ttfamily
  [\href{https://arxiv.org/abs/1704.05871}{arXiv:1704.05871}]}. [Erratum:
  Phys.Rev.D 101, 089902 (2020)].

\bibitem{PhysRevD.101.089902}
M.~Hindmarsh, S.~J.~Huber, K.~Rummukainen, and D.~J.~Weir, ``Erratum: Shape of
  the acoustic gravitational wave power spectrum from a first order phase
  transition [Phys. Rev. D 96, 103520 (2017)],''
  \href{https://dx.doi.org/10.1103/PhysRevD.101.089902}{Phys.\  Rev.\  D
  {\bfseries 101} (2020) 089902}.

\bibitem{Espinosa:2010hh}
J.~R.~Espinosa, T.~Konstandin, J.~M.~No, and G.~Servant, ``{Energy Budget of
  Cosmological First-order Phase Transitions},''
  \href{https://dx.doi.org/10.1088/1475-7516/2010/06/028}{JCAP {\bfseries 06}
  (2010) 028} {\ttfamily
  [\href{https://arxiv.org/abs/1004.4187}{arXiv:1004.4187}]}.

\bibitem{Yagi:2011wg}
K.~Yagi and N.~Seto, ``{Detector configuration of DECIGO/BBO and identification
  of cosmological neutron-star binaries},''
  \href{https://dx.doi.org/10.1103/PhysRevD.83.044011}{Phys.\  Rev.\  D
  {\bfseries 83} (2011) 044011} {\ttfamily
  [\href{https://arxiv.org/abs/1101.3940}{arXiv:1101.3940}]}. [Erratum:
  Phys.Rev.D 95, 109901 (2017)].

\bibitem{Klein:2015hvg}
A.~Klein \emph{et al}., ``{Science with the space-based interferometer eLISA:
  Supermassive black hole binaries},''
  \href{https://dx.doi.org/10.1103/PhysRevD.93.024003}{Phys.\  Rev.\  D
  {\bfseries 93} (2016) 024003} {\ttfamily
  [\href{https://arxiv.org/abs/1511.05581}{arXiv:1511.05581}]}.

\bibitem{Ellis:2020ivx}
S.~A.~R.~Ellis, J.~Quevillon, P.~N.~H.~Vuong, T.~You, and Z.~Zhang, ``{The
  Fermionic Universal One-Loop Effective Action},''
  \href{https://dx.doi.org/10.1007/JHEP11(2020)078}{JHEP {\bfseries 11} (2020)
  078} {\ttfamily [\href{https://arxiv.org/abs/2006.16260}{arXiv:2006.16260}]}.

\bibitem{Coleman:1973jx}
S.~R.~Coleman and E.~J.~Weinberg, ``{Radiative Corrections as the Origin of
  Spontaneous Symmetry Breaking},''
  \href{https://dx.doi.org/10.1103/PhysRevD.7.1888}{Phys.\  Rev.\  D {\bfseries
  7} (1973) 1888--1910}.

\bibitem{PhysRevD.9.3320}
L.~Dolan and R.~Jackiw, ``Symmetry behavior at finite temperature,''
  \href{https://dx.doi.org/10.1103/PhysRevD.9.3320}{Phys.\  Rev.\  D {\bfseries
  9} (1974) 3320--3341}.

\bibitem{PhysRevD.45.2933}
M.~E.~Carrington, ``Effective potential at finite temperature in the standard
  model,'' \href{https://dx.doi.org/10.1103/PhysRevD.45.2933}{Phys.\  Rev.\  D
  {\bfseries 45} (1992) 2933--2944}.

\bibitem{Ellis:2018mja}
J.~Ellis, M.~Lewicki, and J.~M.~No, ``{On the Maximal Strength of a First-Order
  Electroweak Phase Transition and its Gravitational Wave Signal},''
  \href{https://dx.doi.org/10.1088/1475-7516/2019/04/003}{JCAP {\bfseries 04}
  (2019) 003} {\ttfamily
  [\href{https://arxiv.org/abs/1809.08242}{arXiv:1809.08242}]}.

\bibitem{Masoumi:2016wot}
A.~Masoumi, K.~D.~Olum, and B.~Shlaer, ``{Efficient numerical solution to
  vacuum decay with many fields},''
  \href{https://dx.doi.org/10.1088/1475-7516/2017/01/051}{JCAP {\bfseries 01}
  (2017) 051} {\ttfamily
  [\href{https://arxiv.org/abs/1610.06594}{arXiv:1610.06594}]}.

\bibitem{DeBlas:2019qco}
J.~De~Blas, G.~Durieux, C.~Grojean, J.~Gu, and A.~Paul, ``{On the future of
  Higgs, electroweak and diboson measurements at lepton colliders},''
  \href{https://dx.doi.org/10.1007/JHEP12(2019)117}{JHEP {\bfseries 12} (2019)
  117} {\ttfamily [\href{https://arxiv.org/abs/1907.04311}{arXiv:1907.04311}]}.

\bibitem{deBlas:2022ofj}
J.~de~Blas, Y.~Du, C.~Grojean, J.~Gu, \emph{et al}., ``{Global SMEFT Fits at
  Future Colliders},'' in {\em {2022 Snowmass Summer Study}}.
\newblock 2022.
\newblock {\ttfamily
  \href{https://arxiv.org/abs/2206.08326}{arXiv:2206.08326}}.

\bibitem{Ethier:2021bye}
{\bfseries SMEFiT} Collaboration, ``{Combined SMEFT interpretation of Higgs,
  diboson, and top quark data from the LHC},''
  \href{https://dx.doi.org/10.1007/JHEP11(2021)089}{JHEP {\bfseries 11} (2021)
  089} {\ttfamily [\href{https://arxiv.org/abs/2105.00006}{arXiv:2105.00006}]}.

\end{thebibliography}\endgroup

\end{document}